\documentclass[twocolumn,preprintnumbers,amsmath,amssymb,aps,prb,reprint,longbibliography]{revtex4-1}
\usepackage{graphicx}
\begin{document}

\title{
Melting, Reentrant Ordering and Peak Effect For Wigner Crystals with Quenched
and Thermal Disorder
} 
\author{
C. Reichhardt and C. J. O. Reichhardt 
} 
\affiliation{
Theoretical Division and Center for Nonlinear Studies,
Los Alamos National Laboratory, Los Alamos, New Mexico 87545, USA\\ 
} 

\date{\today}
\begin{abstract}
We consider simulations of Wigner crystals interacting with random quenched disorder in the presence of thermal fluctuations. When quenched disorder is absent, there is a well defined melting temperature determined by the proliferation of topological defects, while for zero temperature, there is a critical quenched disorder strength above which topological defects proliferate. When both thermal and quenched disorder are present, these effects compete, and the thermal fluctuations can reduce the effectiveness of the quenched disorder, leading to a reentrant ordered phase in agreement with the predictions of Nelson [Phys. Rev. B 27, 2902 (1983)]. The onset of the reentrant phase can be deduced based on changes in the transport response, where the reentrant ordering appears as an increase in the mobility or the occurrence of a depinning transition. We also find that when the system is in the ordered state and thermally melts, there is an increase in the effective damping or pinning. This produces a drop in the electron mobility that is similar to the peak effect phenomenon found in superconducting vortices, where thermal effects soften the lattice or break down its elasticity, allowing the particles to better adjust their positions to take full advantage of the quenched disorder. 
\end{abstract}

\maketitle
There is a wide class of
systems of two-dimensional (2D) particle assemblies
that form triangular crystalline 
phases, including vortices in thin film superconductors \cite{Guillamon09},
colloidal particles \cite{Murray87,Zahn99,Reichhardt03},
dusty plasmas \cite{Thomas94,Chiang96},
magnetic skyrmions \cite{Huang20,Zazvorka20}, 
active matter \cite{Digregorio22},
and electron solids or Wigner crystals \cite{Monceau12,Shayegan22}. 
In the absence of quenched disorder, these systems exhibit a
melting transition under increasing temperature
that can be characterized 
by the proliferation of topological defects such 
as dislocations and disclinations \cite{Zahn99,Strandburg88,vonGrunberg04}. 
The melting transition in 2D systems has been intensely studied 
and can occur as a two step transition with an
intermediate hexatic phase or
as a single weakly first order phase transition 
\cite{Strandburg88,Marcus96,Du17}.
Even if a hexatic phase is present, the two steps of the transition
can occur so close together that it becomes
difficult to distinguish whether there are two separate phases
in addition to the solid phase \cite{Zahn99}.

Crystalline systems confined to 2D can also show order to disorder transitions 
as a function of increasing quenched disorder 
strength.
At $T = 0$ there can be a  critical
disorder strength above which an amorphous phase appears
\cite{Cha95,Reichhardt17}.
When there are both quenched disorder
and thermal fluctuations,
there can be a competition in which
thermal fluctuations soften the interactions between particles
but at the same time also reduce the
effectiveness of the disordered substrate.
Nelson \cite{Nelson83}
proposed that for a 2D system
with quenched disorder, thermal effects can wash out the
quenched disorder,
leading to a thermally induced reentrant ordering effect 
from a low temperature amorphous state to a crystal state as a function
of increasing temperature. At still higher temperatures,
the system thermally melts.
In other studies, it was suggested
that the quenched disorder always enhances the appearance of
a disordered state,
and that introduction of quenched disorder
monotonically decreases the temperature at which the
transition from a disorder-induced amorphous state to a thermally
reordered state occurs
\cite{Cha95,Reichhardt17,Deutschlander13}.

\begin{figure}
\includegraphics[width=\columnwidth]{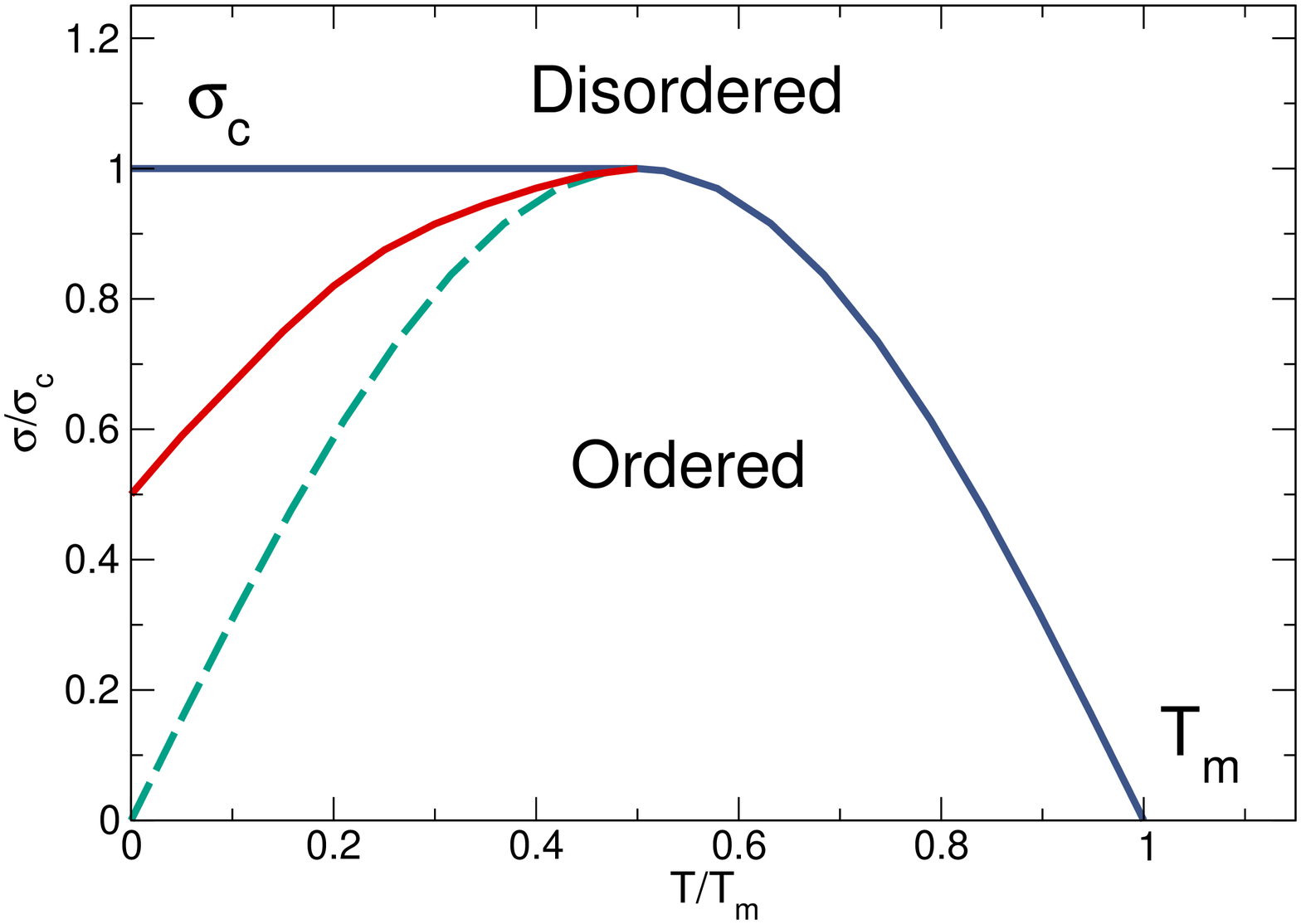}
\caption{
Schematic phase diagram as a function of
reduced quenched disorder strength $\sigma/\sigma_c$ versus
reduced temperature $T/T_m$ for a 2D assembly of repulsive particles in the presence
of quenched disorder.
Here $\sigma_c$ is the quenched disorder strength at the threshold for
dislocation proliferation at $T=0$, while $T_m$ is the melting
temperature at $\sigma=0$.
The green dashed line is the prediction
from Nelson \cite{Nelson83} showing a reentrant disordered phase.
The solid blue line is the prediction from Cha and Fertig \cite{Cha95},
and the solid red line is the result we find in this work. }
\label{fig:1}
\end{figure}

It is possible that the
reentrant ordering proposed by Nelson \cite{Nelson83}
may depend on the size scale and strength of the
quenched disorder pinning sites.
For systems with long range particle-particle interactions but short range
particle-pin interactions, reentrant ordering is possible when
the pinning sites are sufficiently small.
In Fig.~\ref{fig:1} we show a schematic
phase diagram as a function of
reduced disorder strength $\sigma/\sigma_c$ 
versus temperature $T/T_m$ highlighting the ordered and disordered phases. 
When $\sigma = 0$, a melting transition
occurs at $T_m$.
The schematic does not 
distinguish whether there is also a hexatic
phase, but simply indicates the point at which
topological defects start to proliferate.
At $T = 0$, for
increasing $\sigma$ there is a transition from
a crystal to a
disordered state at $\sigma_{c}$.
On the right hand side of the figure, the solid line indicates that the
critical value $\sigma_c$ at which the system disorders decreases with
increasing temperature.
The three lines on the left hand side of the figure indicate different
possible low temperature behaviors.
The dashed line
is the prediction from Nelson \cite{Nelson83}, where the system is disordered
at very low $T$ but thermal effects wash out the effect of the disorder,
permitting reordering to occur with increasing temperature.
Cha and Fertig \cite{Cha95} predicted the upper blue solid line,
where the system remains ordered at low temperatures
all the way up to a constant critical $\sigma_{c}$.  
The solid red line is what we observe in the present work,
where disordering occurs
even at $T=0$ when $\sigma$ is increased, but 
for $\sigma < \sigma_{c}$ 
there is a reentrant ordering with increasing temperature,
so that the predictions of both Nelson and of Cha and Fertig occur.
We also find that for finite $\sigma$,
the thermally induced melting transition at higher temperatures is
depressed in temperature.
We note that in their simulation work,
Cha and Fertig found that at $T = 0$ there is
a critical disorder strength  $\sigma_{c}$
for the 2D crystal to disorder;
however, they did not consider the
effect of finite temperature to see whether there might still be a
reentrant thermally reordered phase.

Anther feature of 2D systems with quenched disorder is
that the transport of the system 
should strongly depend
on whether the particle arrangement is crystalline, disordered,
or fluid
\cite{Reichhardt17,Reichhardt02,Pertsinidis08,Fily10,DiScala12,Reichhardt15}.
If the system is 
in a crystal state, there can still be a depinning threshold,
but the depinning will be an elastic process in which
particles keep their same neighbors
\cite{Reichhardt17,Reichhardt02,DiScala12}. If the 
system is disordered or glassy,
the depinning can be plastic where a portion of particles
remain immobile while the other particles move,
creating local tearing in the assembly
\cite{Reichhardt17,Reichhardt02,Fily10}.
In the fluid phase, strong thermal hopping
reduces the effects of the quenched disorder.
For $T = 0$, the depinning threshold $F_c$
shows a pronounced change across $\sigma_{c}$,
where in general
$F_{c}$ 
rapidly increases with increasing $\sigma$
once the system is on the disordered side of the transition
where the depinning is plastic \cite{Reichhardt17,Reichhardt02,DiScala12}.
Additionally, both the shape of the velocity-force
curves and the fluctuations in the moving
state show pronounced changes across
the transition from elastic to plastic depinning,
going from a single step depinning process on the elastic
side to a multiple step process in the plastic phase
\cite{Reichhardt17,Reichhardt02,Pertsinidis08,Fily10,DiScala12}. 
The jump up in pinning effectiveness upon crossing from elastic
to plastic depinning
has previously been argued to be the cause of
what is called the peak effect
phenomenon observed in type-II
superconductors
\cite{Pippard69,Kes83,Wordenweber86a,Bhattacharya93,Kwok94,Gammel98,Banerjee99,Paltiel00a,Ling01,Troyanovski02,Hilke03,Pasquini08,Okuma08}.
In the superconducting vortex
system, the driving forces on the vortices arise from an
applied current $J$, and the vortices
depin above a critical current $J_{c}$.
The peak effect occurs when,
as a function of increasing temperature, $J_{c}$ undergoes a rapid
increase at a well defined temperature or
magnetic field. If the vortices are already moving when the temperature
is increased, there is a drop in their
average velocity at the peak effect temperature. 
The apparent increase in the effectiveness of the pinning with
increasing temperature
is counterintuitive because it is more natural to expect
that increasing the temperature would reduce the effectiveness of the
pinning.
In the peak effect regime, the thermal fluctuations
are argued to reduce the elasticity 
of the vortex lattice (VL) or induce the formation of dislocations that
strongly soften the VL, causing the VL to
become amorphous and permitting individual vortices
to easily adjust to the pinning landscape
in order to become better pinned \cite{Pippard69,Bhattacharya93,Kwok94}. 
As the temperature is increased further,
the thermal fluctuations overwhelm the pinning energy
and the vortices easily hop out of the pinning sites, causing a decrease in
the critical current or equivalently an increase in the vortex velocity
at fixed current.
It is also possible to observe the peak effect
as a function of increasing field,
and in this case it is argued that
changes in the magnetic penetration depth can reduce the 
strength of the vortex-vortex interactions and soften the VL.
Various measures of the VL structure show that the peak effect is 
characterized by a transition from
an ordered lattice to a disordered or amorphous state
\cite{Gammel98,Ling01,Troyanovski02}.   
The peak effect occurs in both 2D and three-dimensional (3D) systems,
and the 3D peak effect is associated with first order characteristics and
history dependence
\cite{Ling01,Pasquini08}. The peak effect
should be a general feature in any type of 2D elastic system
coupled to quenched disorder where thermal effects
can cause a softening of the lattice, leading to
an increase in the depinning threshold or a drop in the
velocity across the elastic to plastic transition.

Another system that forms a 2D crystal state that can be driven 
is electron solids or Wigner 
crystals
\cite{Wigner34,Andrei88,Goldman90,Jiang91,Williams91,Piacente05,Jang17,Brussarski18,Hatke19,Hossain22}.
These systems usually contain some form 
of quenched disorder, and a variety of studies have
revealed nonlinear transport and possible depinning thresholds
\cite{Goldman90,Jiang91,Williams91,Piacente05}
associated with enhanced noise \cite{Brussarski18}. 
Wigner crystals can also undergo
melting transitions as a function of increasing temperature
\cite{Chen06,Knighton18,Deng19,Ma20,Kim22}.
There is a
growing number of systems where Wigner crystals could be realized, including
dichalcogenide monolayers \cite{Smolenski21},
moir{\' e} heterostructures \cite{Xu20,Li21}
bilayer systems \cite{Zhou21},
and Wigner crystals at zero field \cite{Falson22},
while
new advances in materials preparation point to a variety of future experiments
that could be done in which the competition between quenched
disorder and thermal effects could be studied \cite{Shayegan22}.  
Unlike colloidal assemblies or superconducting vortices,
imaging experiments for Wigner systems are difficult,
so the existence of order to disorder or melting transitions must be determined
on the basis of some type of response or transport experiments. 
An open question is what the Wigner crystal phase diagram is
as a function of quenched disorder and temperature,
as illustrated schematically in 
Fig.~\ref{fig:1},
and whether the different phases can be deduced from transport measures. 
Anther question is whether Wigner crystals
can also exhibit a peak effect or
an increase in the effectiveness of the pinning as a function of
increasing temperature similar to what is found
in superconducting vortex systems across
a thermally induced disordering transition. 

In this work we perform simulations of a 2D
localized electron system in the presence of
quenched disorder $\sigma$ 
and thermal disorder $T$.
At $\sigma = 0$, there is a well defined melting temperature $T_{m}$ 
characterized by a proliferation of topological defects,
while for $T = 0$ there is a
well defined critical quenched disorder strength
$\sigma_{c}$
above which topological defects proliferate.
We map out the phase diagram as a function of
$\sigma$ versus $T/T_{m}$
and find that when $\sigma < \sigma_{c}$, an
increase in $T$ can cause the system to thermally order
as predicted by Nelson \cite{Nelson83}.
Additionally, when $\sigma$ is finite,
thermal disordering occurs at temperatures lower than $T_m$.
For sufficiently large $\sigma$, the system is always disordered.
We show that the ordered and disordered phases can be detected using transport
signatures, where we apply a finite driving force and measure the changes in the
average velocity.
When $\sigma < \sigma_{c}$, the
system forms a disordered pinned state at low temperatures,
but at the reentrant ordering transition, a lattice with elastic
behavior emerges and the depinning threshold is 
strongly reduced, leading to a finite velocity of the Wigner crystal. 
At higher temperatures where the system thermally melts,
under driving
the velocity drops above the melting transition because
the electrons can partially adjust their positions in order to
maximize their interactions with the quenched disorder,
similar to the drop in vortex mobility seen across
the peak effect as a function of increasing temperature in 
superconducting vortex systems \cite{Pippard69,Kes83,Wordenweber86a,Bhattacharya93,Kwok94,Gammel98,Banerjee99,Paltiel00a,Ling01,Troyanovski02,Hilke03,Pasquini08}.
The peak effect we observe only occurs for $T/T_{m} < 1.0$.
We show that reentrance in the phase diagram
occurs both as a function of increasing disorder strength for fixed
disorder density and as a function of increasing disorder density for
fixed disorder strength,
and that the peak effect remains robust for different values of the drive.

\section{Simulation and System} 
We consider a 2D system with periodic boundary conditions
in the $x$ and $y$ directions containing $N_e$ localized electrons at
an electron density of $n = N_{e}/L^2$, where the system is of size
$L \times L$.
Throughout this work we use
$n=0.44$, $L=36$, and $N_{e} = 572$.
The system also contains
$N_{p}$ localized pinning sites of density $n_p=N_p/L^2$.
The initial electron configuration is obtained via simulated 
annealing,
similar to what was done in previous simulations of Wigner crystals
in the presence of disorder
\cite{Reichhardt01,Reichhardt04,Reichhardt21,Reichhardt22}. 
The equation of motion for electron $i$ in the Wigner crystal is
\begin{equation} 
  \alpha_d {\bf v}_{i} = \sum^{N}_{j}\nabla U(r_{ij}) +  {\bf F}_{p} +
         {\bf F}^{T}_{i}
        + {\bf F}_{D} 
        \ .
\end{equation}
The damping term is $\alpha_{d}$ and the electron-electron repulsive 
interaction potential
is $U(r_{ij}) = q/r_{ij}$,
where $q$ is the electron charge,
${\bf r}_i$ and ${\bf r}_j$ are the positions of electrons $i$ and $j$,
and $r_{ij}=|{\bf r}_i-{\bf r}_j|$.
Since the interactions are of long range, we employ 
a real space version of a modified Ewald summation technique
called the Lekner method as in previous work
\cite{Lekner91,GronbechJensen97a}. 
The pinning force ${\bf F}_{p}$ is modeled as arising from
randomly placed short range parabolic traps
of radius $r_{p}=0.35$. 
The thermal fluctuations ${\bf F}^T$ are represented by Langevin kicks
with the properties
$\langle {\bf F}_i^{T}\rangle = 0$
and $\langle {\bf F}^T_i(t){\bf F}_j^{T}(t^\prime)\rangle = 2k_BT\delta_{ij}\delta(t-t^\prime)$.  
We also consider the effect of an applied driving force
${\bf F}_{D}=F_D{\bf \hat x}$, which could come from an applied voltage. 
We measure the average velocity per particle,
$\langle V\rangle = \sum^{N_e}_i{\bf v}_i\cdot {\hat {\bf x}}$,
allowing us to construct the equivalent of an experimental
current-voltage curve.
The simulation method we employ for the pinning and dynamics of
Wigner crystals was used previously to examine
nonlinear velocity-force curves \cite{Reichhardt01},
noise \cite{Reichhardt04}, and depinning thresholds
\cite{Reichhardt22,Cha94a}. 
If the electrons are subjected to a magnetic field ${\bf B}$, there can be
an additional force term $q{\bf B}\times {\bf v}_{i}$ that can
generate a Hall angle for the electron motion
\cite{Reichhardt21}; however,
in general this term is small and we will not consider it in this work.

\begin{figure}
\includegraphics[width=\columnwidth]{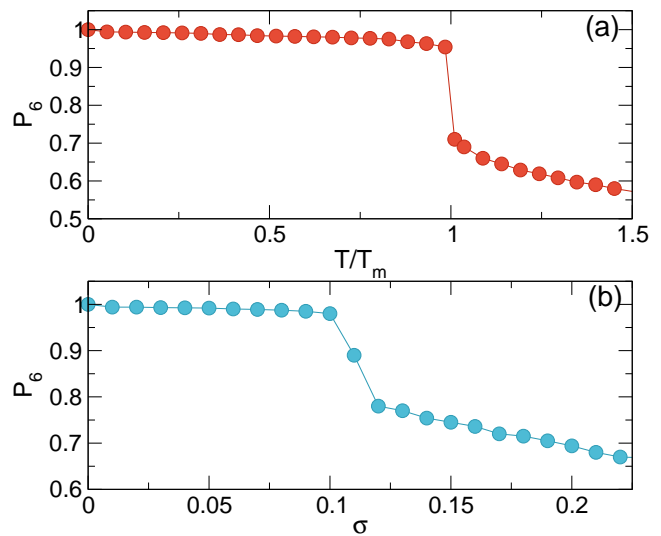}
\caption{(a) The fraction of sixfold coordinated electrons $P_{6}$
vs reduced temperature $T/T_{m}$ for a
system with
no quenched disorder,
$\sigma = 0$. There is a well
defined melting temperature $T_{m}$ indicated by the drop in $P_{6}$.
(b) $P_{6}$ vs disorder strength $\sigma$ in a system with
pinning density $n_p=0.25$ and zero temperature,
$T/T_m= 0$.
There is a well defined
value of $\sigma$, $\sigma_c$,
at which topological defects begin to proliferate,
leading to 
a drop in $P_{6}$.} 
\label{fig:2}
\end{figure}

\begin{figure}
\includegraphics[width=\columnwidth]{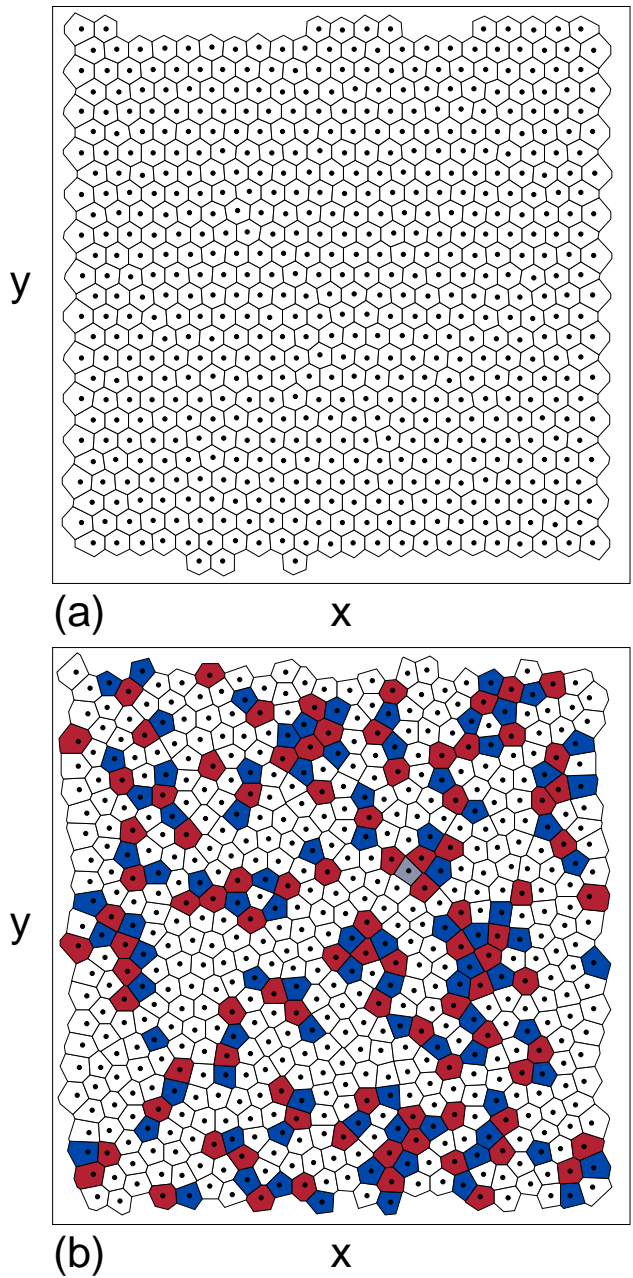}
\caption{Voronoi construction of the electron positions for the system
in Fig.~\ref{fig:2}(a) with
no quenched disorder,
$\sigma=0$. White polygons are sixfold coordinated, blue are fivefold
coordinated, red are sevenfold coordinated, and gray are fourfold coordinated.
(a) At $T/T_{m} = 0.52$ the lattice is triangular.
(b) At $T/T_{m} = 1.09$,
numerous topological defects
have appeared and the system is in a liquid state.} 
\label{fig:3}
\end{figure}

\section{Results}
We first study a sample containing no quenched disorder.
To characterize the system we use the fraction of sixfold
coordinated electrons,
$P_6=N_e^{-1}\sum_i^{N_e} \delta(z_i-6)$, where the coordination
number $z_i$ of electron $i$ is obtained using 
a Voronoi construction.
For a triangular lattice of electrons, $P_{6} = 1.0$.
In Fig.~\ref{fig:2}(a) we  
plot $P_{6}$ versus $T/T_{m}$, where
the melting temperature $T_{m}$ is defined to occur at the point where
$P_{6}$ shows a rapid drop.
Figure~\ref{fig:2}(a) indicates that the melting temperature is
well defined.
We plot the Voronoi construction for the electron positions
at $T/T_m=0.52$ in Fig.~\ref{fig:3}(a), where the lattice is
ordered, and at $T/T_m=1.09$ in
Fig.~\ref{fig:3}(b), where numerous topological defects have appeared.
In Fig.~\ref{fig:2}(b) we show $P_{6}$ versus the maximum disorder strength
$\sigma$ in a sample with a
a pinning density of $n_{p} = 0.25$
at zero temperature, $T=0$.
There is a well defined disorder strength
$\sigma_{c} \approx 0.105$
above which a proliferation of topological defects occurs.

\begin{figure}
\includegraphics[width=\columnwidth]{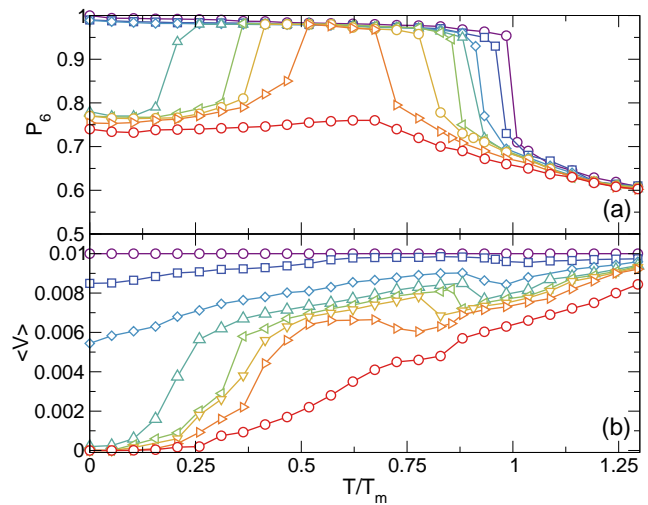}
\caption{(a) $P_{6}$ vs $T/T_{m}$ for the system in
Fig.~\ref{fig:2} with
$n_p=0.25$ at
$\sigma = 0.0$, 0.06, 0.1, 0.12, 0.13, 0.135, 0.14, and $0.16$,
from top to bottom. 
For $\sigma = 0.0$, 0.06, and $0.1$, there is no reentrant ordering,
but for $\sigma=0.12$, 0.13, 0.135, and $0.14$,
there is reentrant ordering with increasing temperature.
(b) The corresponding velocity
$\langle V\rangle$ vs $T/T_{m}$ at $F_{D} = 0.01$.
The velocities are low in the disordered regime at lower $T/T_m$,
but increase at the reentrant ordering
transition.
In addition,
at the thermal melting transition
there is a drop in $\langle V\rangle$ similar to the
peak effect phenomenon found for superconducting vortices.}
\label{fig:4}
\end{figure}

We next perform a series of simulations at different combinations of
$\sigma$ and $T$ for the system from Fig.~\ref{fig:2} with
$n_p=0.25$.
In Fig.~\ref{fig:4}(a) we plot
$P_{6}$ versus $T/T_m$
at 
$\sigma = 0$, 0.06, 0.1, 0.12, 0.13, 0.135, 0.14, and $0.16$,
where $T_m$ is defined to be the temperature
at which melting occurs
for $\sigma= 0$.
For $\sigma = 0.0$, 0.06, and $0.1$,
the system starts off ordered at low $T/T_m$
and remains ordered until it melts near $T/T_m=1$.
The drop in $P_6$ associated with melting shifts to lower values of
$T/T_m$ as $\sigma$ increases.
For $\sigma = 0.12$, 0.13, 0.135, and $0.14$,
all of which are above the value of
$\sigma = 0.105$ for which
the system disorders at $T = 0$ in Fig.~\ref{fig:2}(b), the 
system is disordered at low temperature.
As $T/T_m$ increases, there 
is a critical temperature at which
$P_{6}$ increases back to a value near $P_6=1$,
indicating that the system has ordered under increasing temperature.
The temperature at which this reentrant ordering occurs
shifts to higher $T/T_m$ with increasing $\sigma$,
while the thermal disordering temperature drops to lower $T/T_m$ as $\sigma$
becomes larger, so for
$\sigma = 0.14$ there is only a narrow window,
$0.5 < T/T_m < 0.75$, where the system is ordered. 
For $\sigma = 0.16$, the system is disordered at all temperatures.

In Fig.~\ref{fig:4}(b) we plot
$\langle V\rangle$ versus $T/T_{m}$
for the same system in Fig.~\ref{fig:4}(a) where
we have added a driving force with
$F_{D} = 0.01$. Under this drive, the $\sigma = 0$ system
has $\langle V\rangle = 0.01$ for all values of $T/T_{m}$. 
When $\sigma = 0.06$ or $0.01$,
$\langle V\rangle$ starts off at a finite value and increases
with increasing $T/T_m$; 
however, for $\sigma > 0.1$,
at low temperatures where $P_6$ is low,
$\langle V\rangle = 0$.
This indicates that
the system is easily pinned in the low temperature
disordered phase; however, when the system reaches the reentrant
ordered state, it forms an elastic lattice that is less well pinned,
leading to an increase in $\langle V\rangle$.
Anther interesting effect is that at the thermal melting transition,
the drop in $P_{6}$ is also correlated with a drop in
$\langle V\rangle$, 
indicating that the effectiveness of the
pinning increases at the thermally induced
disordering transition.
This behavior is very similar
to the peak effect phenomenon,
where at a finite drive the
average superconducting vortex velocity
can show a drop with increasing temperature when the system
thermally disorders. In the vortex case, it has been
argued that the disordered state is softer
and can therefore better adapt to the pinning landscape.
Figure~\ref{fig:4}(b) shows that
at higher temperatures where the thermal effects start to dominate,
the velocity increases with increasing $T/T_m$.
For $\sigma = 0.06$ 
there is only a weak dip
in $\langle V\rangle$ at the thermal disordering temperature,
and at $\sigma = 0.16$, there is no dip in the
velocity since the system is always disordered.

\begin{figure}
\includegraphics[width=\columnwidth]{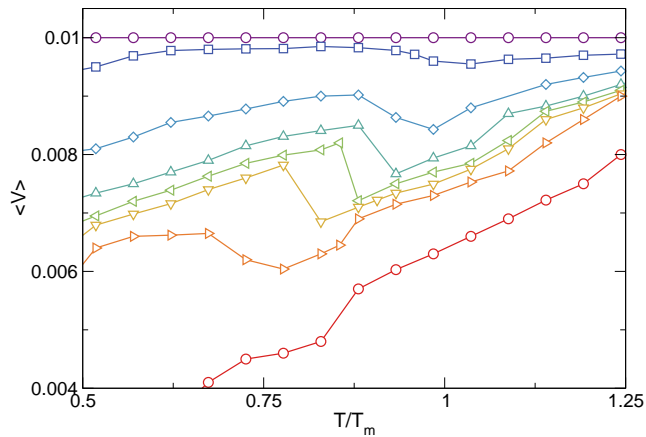}
\caption{A blowup of the plot of $\langle V\rangle$ vs
$T/T_m$ in Fig.~\ref{fig:4}(b) for the system in Fig.~\ref{fig:2} with
$n_p=0.25$ at
$\sigma = 0.0$, 0.06, 0.1, 0.12, 0.13, 0.135, 0.14, and $0.16$,
from top to bottom, 
showing
that the velocity drops across the thermal disordering
transition.}
\label{fig:5}
\end{figure}

In Fig.~\ref{fig:5} we show a blowup of the $\langle V\rangle$ versus
$T/T_m$ plot from Fig.~\ref{fig:4}(b) in order to highlight the
drop in $\langle V\rangle$
across the thermal melting transition. 
In general, the drop in velocity 
only appears at $T/T_{m} < 1.0$ and shifts to lower values
of $T/T_m$ as the pinning strength $\sigma$ increases.
This result indicates
that both the reentrant ordering transition
and the thermal melting can be detected using
changes in the transport measures. 

\begin{figure}
\includegraphics[width=\columnwidth]{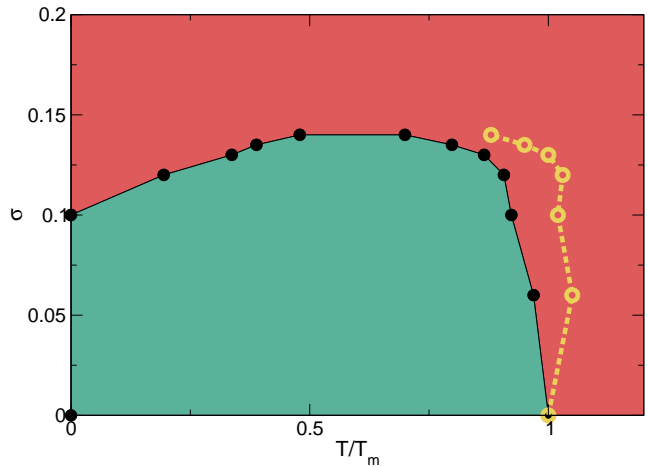}
\caption{
Phase diagram as a function of pinning strength $\sigma$ vs reduced
temperature $T/T_m$ constructed from the features in Fig.~\ref{fig:4} for
a system with
$n_p=0.25$.
Green indicates the ordered regime and red indicates the disordered
regime.
Reentrant ordering occurs for $0.1 < \sigma  < 0.14$.
The peak effect appears between the disordering transition and the
thermal melting regime, and the dashed line indicates the end of the
peak effect window.
}
\label{fig:6}
\end{figure}

Using the features in Fig.~\ref{fig:4},
in Fig.~\ref{fig:6} we construct a phase
diagram
as a function of $\sigma$ versus $T/T_m$ 
showing
the ordered and disordered
regimes.
Reentrant ordering occurs for $0.1 < \sigma < 0.14$,
and the thermally induced disordering transition drops further below
$T/T_{m} = 1.0$ with increasing $\sigma$.
The phase diagram includes the reentrant ordering feature predicted
by Nelson \cite{Nelson83}
as well as a finite value of $\sigma_c$ 
at $T = 0$ as proposed by Cha and Fertig \cite{Cha95}.
For $\sigma > 0.14$, the system is always disordered.
The dashed line indicates the upper end of the region in which there
is a velocity reduction associated with the peak effect.
This line is determined by the point at which $\langle V\rangle$ reaches
75\% of $\langle V_d\rangle$, where $\langle V_d\rangle$ is
the value of $\langle V\rangle$ just before the velocity dip begins as
a function of increasing $T/T_m$.
We note that additional lines could be
drawn in the disordered regime to differentiate
between a low temperature glassy or pinned phase and
a higher temperature fluctuating fluid phase.

\begin{figure}
\includegraphics[width=\columnwidth]{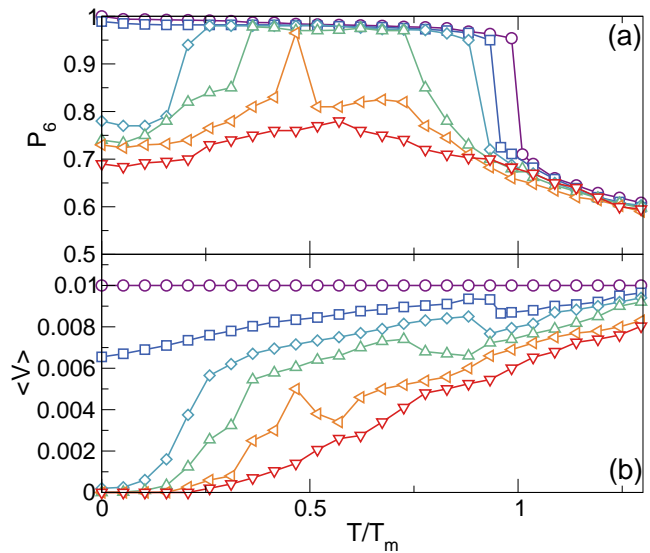}
\caption{(a) $P_{6}$ vs $T/T_{m}$ for the system in
Fig.~\ref{fig:4}
at fixed $\sigma = 0.12$ for  
$n_{p} = 0$, 0.15, 0.25, 0.35, 0.45, and $0.55$, from top  to bottom.  
(b) The corresponding $\langle V\rangle$ vs $T/T_{m}$
under a drive of $F_{D} = 0.01$ showing a drop
in velocity across the thermal melting transition.}
\label{fig:7}
\end{figure}

In Fig.~\ref{fig:7}(a) we plot $P_{6}$
versus $T/T_{m}$  for the system in
Fig.~\ref{fig:4} at a fixed
$\sigma = 0.12$ for
varied pinning densities of $n_{p} = 0$, 0.15, 0.25, 0.35, 0.45 and $0.55$.
For $n_{p} = 0$ and $0.15$ there is no disordered phase
at low $T/T_m$;
however, for $0.25 \leq n_{p} \leq 0.45$,
there is a regime of reentrant ordering. 
For $n_{p} = 0.55$, the system is disordered for all values of $T/T_{m}$.
Figure~\ref{fig:7}(b) shows the
corresponding $\langle V\rangle$ versus $T/T_m$ curves.
The drop in $\langle V\rangle$ at the thermal ordering transition
indicates that the same peak effect appears that was found
for varied disorder strengths.
For $n_{p} = 0.55$, $\langle V\rangle$
monotonically increases with increasing $T/T_m$. 

\begin{figure}
\includegraphics[width=\columnwidth]{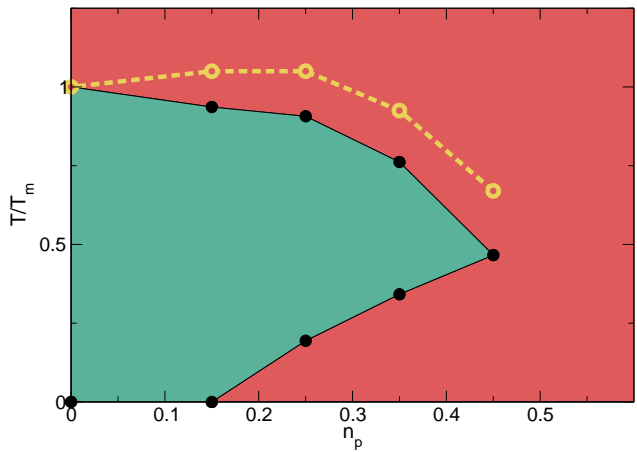}
\caption{
Phase diagram as a function of reduced temperature $T/T_m$ vs
pinning density $n_p$ constructed from the features in Fig.~\ref{fig:7}
for a system with
$\sigma=0.12$.
Green indicates the ordered regime and red indicates the disordered
regime.
Reentrant ordering occurs for
$0.15 < n_{p} < 0.45$.
The dashed line indicates the end of the peak effect window where the
velocity drop is lost.} 
\label{fig:8}
\end{figure}

In Fig.~\ref{fig:8} we show a phase diagram of the ordered and disordered
states as a function of $T/T_m$ versus pinning density $n_p$ constructed
using the features in Fig.~\ref{fig:7}.
The same reentrant ordering
phase appears that was found in Fig.~\ref{fig:6} at
constant pinning density for changing pinning strength.
The dashed
line indicates where the drop in velocity associated with peak effect is lost. 

\begin{figure}
\includegraphics[width=\columnwidth]{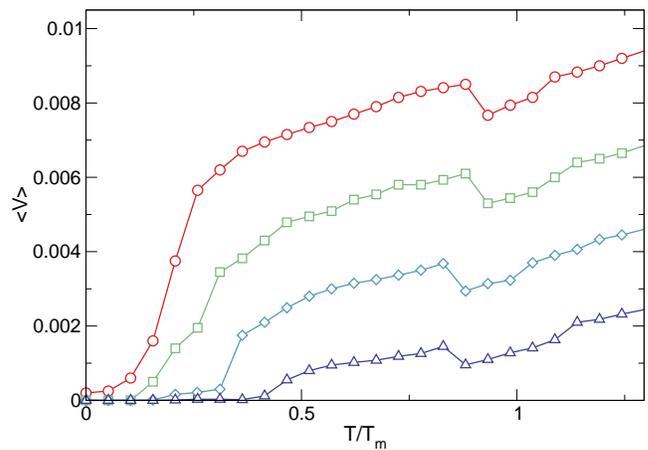}
\caption{
$\langle V\rangle$ vs $T/T_{m}$ curves for a system with $\sigma = 0.12$
and $n_{p} = 0.25$ at $F_{D} = 0.01$, 0.0075, 0.005, and $0.0025$, from
top to bottom.}
\label{fig:9}
\end{figure}

We have also tested how robust the peak effect phenomenon
is for varied driving force $F_{D}$. 
In Fig.~\ref{fig:9} we plot $\langle V\rangle$
versus $T/T_{m}$ for a system with $n_{p}= 0.25$ and $\sigma = 0.12$ 
at $F_{D} = 0.01$, 0.0075, 0.005, and $0.0025$.
As $F_{D}$ decreases,
at low temperatures there is an extended window
where the system is pinned,
while at higher temperatures
the peak effect region remains robust.
The thermal melting transition shifts to slightly lower values
of $T/T_m$ with decreasing driving force.
This result suggests that if $F_{D}$ is low enough, the system 
could show a reentrant pinning near the thermal disordering
transition. 

\section{Summary} 
We have investigated the quenched disorder versus temperature phase diagram for
a 2D Wigner crystal system. 
For zero quenched disorder, there is a
well defined melting temperature that
is characterized by the proliferation of topological defects.
For zero temperature there is also a well defined quenched disorder
strength at which the system becomes disordered.
The phase diagram shows that when the disorder strength is larger than
the value at which the zero temperature system becomes disordered,
as the temperature is increased the system can show a 
thermally induced or reentrant ordering transition
when the 
thermal fluctuations become strong enough to overwhelm the effect of
the quenched disorder without melting the lattice.
At higher temperatures, introduction of quenched disorder reduces the
thermal disordering temperature of the system,
while for strong enough quenched disorder the system is always disordered.
The phase diagrams we obtain show both the reentrant ordering
feature predicted by Nelson for 2D systems with quenched and thermal disorder
as well as a well defined quenched disorder strength where the electrons
disorder at zero temperature, as predicted by Cha and Fertig.
We also show that these phases can be observed
through features in the transport curves,
where the effectiveness of the
pinning strongly drops at the reentrant ordering
transition as the
system goes from plastic behavior to elastic behavior.
In the presence of quenched disorder, the thermal disordering transition
is associated with a drop in 
the average electron velocity,
similar to the peak effect phenomenon
found in superconducting vortex systems.
The drop occurs when thermal fluctuations break down
the elasticity of the crystal
and allow the electrons to adjust more easily to the pinning substrate. 
At higher temperatures, the velocity goes back up again when the electrons
begin to hop readily out of the pinning barriers.
We show that these effects are robust for
a range of pinning strengths, densities, and drives.
Our predictions could be tested by examining transport
signatures in Wigner crystals where both
thermal and quenched disorder effects arise. Our results
should be general to
the broader class of 2D systems with quenched disorder.

\begin{acknowledgments}
We gratefully acknowledge the support of the U.S. Department of
Energy through the LANL/LDRD program for this work.
This work was supported by the US Department of Energy through
the Los Alamos National Laboratory.  Los Alamos National Laboratory is
operated by Triad National Security, LLC, for the National Nuclear Security
Administration of the U. S. Department of Energy (Contract No. 892333218NCA000001).

\end{acknowledgments}

\bibliography{mybib}

\begin{thebibliography}{64}%
\makeatletter
\providecommand \@ifxundefined [1]{%
 \@ifx{#1\undefined}
}%
\providecommand \@ifnum [1]{%
 \ifnum #1\expandafter \@firstoftwo
 \else \expandafter \@secondoftwo
 \fi
}%
\providecommand \@ifx [1]{%
 \ifx #1\expandafter \@firstoftwo
 \else \expandafter \@secondoftwo
 \fi
}%
\providecommand \natexlab [1]{#1}%
\providecommand \enquote  [1]{``#1''}%
\providecommand \bibnamefont  [1]{#1}%
\providecommand \bibfnamefont [1]{#1}%
\providecommand \citenamefont [1]{#1}%
\providecommand \href@noop [0]{\@secondoftwo}%
\providecommand \href [0]{\begingroup \@sanitize@url \@href}%
\providecommand \@href[1]{\@@startlink{#1}\@@href}%
\providecommand \@@href[1]{\endgroup#1\@@endlink}%
\providecommand \@sanitize@url [0]{\catcode `\\12\catcode `\$12\catcode
  `\&12\catcode `\#12\catcode `\^12\catcode `\_12\catcode `\%12\relax}%
\providecommand \@@startlink[1]{}%
\providecommand \@@endlink[0]{}%
\providecommand \url  [0]{\begingroup\@sanitize@url \@url }%
\providecommand \@url [1]{\endgroup\@href {#1}{\urlprefix }}%
\providecommand \urlprefix  [0]{URL }%
\providecommand \Eprint [0]{\href }%
\providecommand \doibase [0]{http://dx.doi.org/}%
\providecommand \selectlanguage [0]{\@gobble}%
\providecommand \bibinfo  [0]{\@secondoftwo}%
\providecommand \bibfield  [0]{\@secondoftwo}%
\providecommand \translation [1]{[#1]}%
\providecommand \BibitemOpen [0]{}%
\providecommand \bibitemStop [0]{}%
\providecommand \bibitemNoStop [0]{.\EOS\space}%
\providecommand \EOS [0]{\spacefactor3000\relax}%
\providecommand \BibitemShut  [1]{\csname bibitem#1\endcsname}%
\let\auto@bib@innerbib\@empty
\bibitem [{\citenamefont {Guillamon}\ \emph {et~al.}(2009)\citenamefont
  {Guillamon}, \citenamefont {Suderow}, \citenamefont {Fernandez-Pacheco},
  \citenamefont {Sese}, \citenamefont {Cordoba}, \citenamefont {De~Teresa},
  \citenamefont {Ibarra},\ and\ \citenamefont {Vieira}}]{Guillamon09}%
  \BibitemOpen
  \bibfield  {author} {\bibinfo {author} {\bibfnamefont {I.}~\bibnamefont
  {Guillamon}}, \bibinfo {author} {\bibfnamefont {H.}~\bibnamefont {Suderow}},
  \bibinfo {author} {\bibfnamefont {A.}~\bibnamefont {Fernandez-Pacheco}},
  \bibinfo {author} {\bibfnamefont {J.}~\bibnamefont {Sese}}, \bibinfo {author}
  {\bibfnamefont {R.}~\bibnamefont {Cordoba}}, \bibinfo {author} {\bibfnamefont
  {J.~M.}\ \bibnamefont {De~Teresa}}, \bibinfo {author} {\bibfnamefont {M.~R.}\
  \bibnamefont {Ibarra}}, \ and\ \bibinfo {author} {\bibfnamefont
  {S.}~\bibnamefont {Vieira}},\ }\bibfield  {title} {\enquote {\bibinfo {title}
  {Direct observation of melting in a two-dimensional superconducting vortex
  lattice},}\ }\href {\doibase 10.1038/NPHYS1368} {\bibfield  {journal}
  {\bibinfo  {journal} {Nature Phys.}\ }\textbf {\bibinfo {volume} {5}},\
  \bibinfo {pages} {651--655} (\bibinfo {year} {2009})}\BibitemShut {NoStop}%
\bibitem [{\citenamefont {Murray}\ and\ \citenamefont
  {Van~Winkle}(1987)}]{Murray87}%
  \BibitemOpen
  \bibfield  {author} {\bibinfo {author} {\bibfnamefont {C.~A.}\ \bibnamefont
  {Murray}}\ and\ \bibinfo {author} {\bibfnamefont {D.~H.}\ \bibnamefont
  {Van~Winkle}},\ }\bibfield  {title} {\enquote {\bibinfo {title} {Experimental
  observation of two-stage melting in a classical two-dimensional screened
  {C}oulomb system},}\ }\href {\doibase 10.1103/PhysRevLett.58.1200} {\bibfield
   {journal} {\bibinfo  {journal} {Phys. Rev. Lett.}\ }\textbf {\bibinfo
  {volume} {58}},\ \bibinfo {pages} {1200--1203} (\bibinfo {year}
  {1987})}\BibitemShut {NoStop}%
\bibitem [{\citenamefont {Zahn}\ \emph {et~al.}(1999)\citenamefont {Zahn},
  \citenamefont {Lenke},\ and\ \citenamefont {Maret}}]{Zahn99}%
  \BibitemOpen
  \bibfield  {author} {\bibinfo {author} {\bibfnamefont {K.}~\bibnamefont
  {Zahn}}, \bibinfo {author} {\bibfnamefont {R.}~\bibnamefont {Lenke}}, \ and\
  \bibinfo {author} {\bibfnamefont {G.}~\bibnamefont {Maret}},\ }\bibfield
  {title} {\enquote {\bibinfo {title} {Two-stage melting of paramagnetic
  colloidal crystals in two dimensions},}\ }\href {\doibase
  10.1103/PhysRevLett.82.2721} {\bibfield  {journal} {\bibinfo  {journal}
  {Phys. Rev. Lett.}\ }\textbf {\bibinfo {volume} {82}},\ \bibinfo {pages}
  {2721--2724} (\bibinfo {year} {1999})}\BibitemShut {NoStop}%
\bibitem [{\citenamefont {Reichhardt}\ and\ \citenamefont
  {Olson~Reichhardt}(2003)}]{Reichhardt03}%
  \BibitemOpen
  \bibfield  {author} {\bibinfo {author} {\bibfnamefont {C.}~\bibnamefont
  {Reichhardt}}\ and\ \bibinfo {author} {\bibfnamefont {C.~J.}\ \bibnamefont
  {Olson~Reichhardt}},\ }\bibfield  {title} {\enquote {\bibinfo {title}
  {Fluctuating topological defects in {2D} liquids: Heterogeneous motion and
  noise},}\ }\href {\doibase 10.1103/PhysRevLett.90.095504} {\bibfield
  {journal} {\bibinfo  {journal} {Phys. Rev. Lett.}\ }\textbf {\bibinfo
  {volume} {90}},\ \bibinfo {pages} {095504} (\bibinfo {year}
  {2003})}\BibitemShut {NoStop}%
\bibitem [{\citenamefont {Thomas}\ \emph {et~al.}(1994)\citenamefont {Thomas},
  \citenamefont {Morfill}, \citenamefont {Demmel}, \citenamefont {Goree},
  \citenamefont {Feuerbacher},\ and\ \citenamefont {M\"ohlmann}}]{Thomas94}%
  \BibitemOpen
  \bibfield  {author} {\bibinfo {author} {\bibfnamefont {H.}~\bibnamefont
  {Thomas}}, \bibinfo {author} {\bibfnamefont {G.~E.}\ \bibnamefont {Morfill}},
  \bibinfo {author} {\bibfnamefont {V.}~\bibnamefont {Demmel}}, \bibinfo
  {author} {\bibfnamefont {J.}~\bibnamefont {Goree}}, \bibinfo {author}
  {\bibfnamefont {B.}~\bibnamefont {Feuerbacher}}, \ and\ \bibinfo {author}
  {\bibfnamefont {D.}~\bibnamefont {M\"ohlmann}},\ }\bibfield  {title}
  {\enquote {\bibinfo {title} {Plasma crystal: {C}oulomb crystallization in a
  dusty plasma},}\ }\href {\doibase 10.1103/PhysRevLett.73.652} {\bibfield
  {journal} {\bibinfo  {journal} {Phys. Rev. Lett.}\ }\textbf {\bibinfo
  {volume} {73}},\ \bibinfo {pages} {652--655} (\bibinfo {year}
  {1994})}\BibitemShut {NoStop}%
\bibitem [{\citenamefont {Chiang}\ and\ \citenamefont {I}(1996)}]{Chiang96}%
  \BibitemOpen
  \bibfield  {author} {\bibinfo {author} {\bibfnamefont {C.-H.}\ \bibnamefont
  {Chiang}}\ and\ \bibinfo {author} {\bibfnamefont {L.}~\bibnamefont {I}},\
  }\bibfield  {title} {\enquote {\bibinfo {title} {Cooperative particle motions
  and dynamical behaviors of free dislocations in strongly coupled quasi-{2D}
  dusty plasmas},}\ }\href {\doibase 10.1103/PhysRevLett.77.647} {\bibfield
  {journal} {\bibinfo  {journal} {Phys. Rev. Lett.}\ }\textbf {\bibinfo
  {volume} {77}},\ \bibinfo {pages} {647--650} (\bibinfo {year}
  {1996})}\BibitemShut {NoStop}%
\bibitem [{\citenamefont {Huang}\ \emph {et~al.}(2020)\citenamefont {Huang},
  \citenamefont {Schonenberger}, \citenamefont {Cantoni}, \citenamefont
  {Heinen}, \citenamefont {Magrez}, \citenamefont {Rosch}, \citenamefont
  {Carbone},\ and\ \citenamefont {R{\o}nnow}}]{Huang20}%
  \BibitemOpen
  \bibfield  {author} {\bibinfo {author} {\bibfnamefont {P.}~\bibnamefont
  {Huang}}, \bibinfo {author} {\bibfnamefont {T.}~\bibnamefont
  {Schonenberger}}, \bibinfo {author} {\bibfnamefont {M.}~\bibnamefont
  {Cantoni}}, \bibinfo {author} {\bibfnamefont {L.}~\bibnamefont {Heinen}},
  \bibinfo {author} {\bibfnamefont {A.}~\bibnamefont {Magrez}}, \bibinfo
  {author} {\bibfnamefont {A.}~\bibnamefont {Rosch}}, \bibinfo {author}
  {\bibfnamefont {F.}~\bibnamefont {Carbone}}, \ and\ \bibinfo {author}
  {\bibfnamefont {H.~M.}\ \bibnamefont {R{\o}nnow}},\ }\bibfield  {title}
  {\enquote {\bibinfo {title} {Melting of a skyrmion lattice to a skyrmion
  liquid via a hexatic phase},}\ }\href {\doibase 10.1038/s41565-020-0716-3}
  {\bibfield  {journal} {\bibinfo  {journal} {Nature Nanotechnol.}\ }\textbf
  {\bibinfo {volume} {15}},\ \bibinfo {pages} {761} (\bibinfo {year}
  {2020})}\BibitemShut {NoStop}%
\bibitem [{\citenamefont {Z{\' a}zvorka}\ \emph {et~al.}(2020)\citenamefont
  {Z{\' a}zvorka}, \citenamefont {Dittrich}, \citenamefont {Ge}, \citenamefont
  {Kerber}, \citenamefont {Raab}, \citenamefont {Winkler}, \citenamefont
  {Litzius}, \citenamefont {Veis}, \citenamefont {Virnau},\ and\ \citenamefont
  {Kl{\" a}ui}}]{Zazvorka20}%
  \BibitemOpen
  \bibfield  {author} {\bibinfo {author} {\bibfnamefont {J.}~\bibnamefont {Z{\'
  a}zvorka}}, \bibinfo {author} {\bibfnamefont {F.}~\bibnamefont {Dittrich}},
  \bibinfo {author} {\bibfnamefont {Y.}~\bibnamefont {Ge}}, \bibinfo {author}
  {\bibfnamefont {N.}~\bibnamefont {Kerber}}, \bibinfo {author} {\bibfnamefont
  {K.}~\bibnamefont {Raab}}, \bibinfo {author} {\bibfnamefont {T.}~\bibnamefont
  {Winkler}}, \bibinfo {author} {\bibfnamefont {K.}~\bibnamefont {Litzius}},
  \bibinfo {author} {\bibfnamefont {M.}~\bibnamefont {Veis}}, \bibinfo {author}
  {\bibfnamefont {P.}~\bibnamefont {Virnau}}, \ and\ \bibinfo {author}
  {\bibfnamefont {M.}~\bibnamefont {Kl{\" a}ui}},\ }\bibfield  {title}
  {\enquote {\bibinfo {title} {Skyrmion lattice phases in thin film
  multilayer},}\ }\href {\doibase 10.1002/adfm.202004037} {\bibfield  {journal}
  {\bibinfo  {journal} {Adv. Funct. Mater.}\ }\textbf {\bibinfo {volume}
  {30}},\ \bibinfo {pages} {2004037} (\bibinfo {year} {2020})}\BibitemShut
  {NoStop}%
\bibitem [{\citenamefont {Digregorio}\ \emph {et~al.}(2022)\citenamefont
  {Digregorio}, \citenamefont {Levis}, \citenamefont {Cugliandolo},
  \citenamefont {Gonnella},\ and\ \citenamefont
  {Pagonabarraga}}]{Digregorio22}%
  \BibitemOpen
  \bibfield  {author} {\bibinfo {author} {\bibfnamefont {P.}~\bibnamefont
  {Digregorio}}, \bibinfo {author} {\bibfnamefont {D.}~\bibnamefont {Levis}},
  \bibinfo {author} {\bibfnamefont {L.~F.}\ \bibnamefont {Cugliandolo}},
  \bibinfo {author} {\bibfnamefont {G.}~\bibnamefont {Gonnella}}, \ and\
  \bibinfo {author} {\bibfnamefont {I.}~\bibnamefont {Pagonabarraga}},\
  }\bibfield  {title} {\enquote {\bibinfo {title} {Unified analysis of
  topological defects in {2D} systems of active and passive disks},}\ }\href
  {\doibase 10.1039/D1SM01411K} {\bibfield  {journal} {\bibinfo  {journal}
  {Soft Matter}\ }\textbf {\bibinfo {volume} {18}},\ \bibinfo {pages}
  {566--591} (\bibinfo {year} {2022})}\BibitemShut {NoStop}%
\bibitem [{\citenamefont {Monceau}(2012)}]{Monceau12}%
  \BibitemOpen
  \bibfield  {author} {\bibinfo {author} {\bibfnamefont {P.}~\bibnamefont
  {Monceau}},\ }\bibfield  {title} {\enquote {\bibinfo {title} {Electronic
  crystals: an experimental overview},}\ }\href {\doibase
  10.1080/00018732.2012.719674} {\bibfield  {journal} {\bibinfo  {journal}
  {Adv. Phys.}\ }\textbf {\bibinfo {volume} {61}},\ \bibinfo {pages} {325--581}
  (\bibinfo {year} {2012})}\BibitemShut {NoStop}%
\bibitem [{\citenamefont {Shayegan}(2022)}]{Shayegan22}%
  \BibitemOpen
  \bibfield  {author} {\bibinfo {author} {\bibfnamefont {M.}~\bibnamefont
  {Shayegan}},\ }\bibfield  {title} {\enquote {\bibinfo {title} {Wigner
  crystals in flat band {2D} electron systems},}\ }\href {\doibase
  10.1038/s42254-022-00444-4} {\bibfield  {journal} {\bibinfo  {journal}
  {Nature Rev. Phys.}\ }\textbf {\bibinfo {volume} {4}},\ \bibinfo {pages}
  {212--213} (\bibinfo {year} {2022})}\BibitemShut {NoStop}%
\bibitem [{\citenamefont {Strandburg}(1988)}]{Strandburg88}%
  \BibitemOpen
  \bibfield  {author} {\bibinfo {author} {\bibfnamefont {K.~J.}\ \bibnamefont
  {Strandburg}},\ }\bibfield  {title} {\enquote {\bibinfo {title}
  {Two-dimensional melting},}\ }\href {\doibase 10.1103/RevModPhys.60.161}
  {\bibfield  {journal} {\bibinfo  {journal} {Rev. Mod. Phys.}\ }\textbf
  {\bibinfo {volume} {60}},\ \bibinfo {pages} {161--207} (\bibinfo {year}
  {1988})}\BibitemShut {NoStop}%
\bibitem [{\citenamefont {von Gr\"unberg}\ \emph {et~al.}(2004)\citenamefont
  {von Gr\"unberg}, \citenamefont {Keim}, \citenamefont {Zahn},\ and\
  \citenamefont {Maret}}]{vonGrunberg04}%
  \BibitemOpen
  \bibfield  {author} {\bibinfo {author} {\bibfnamefont {H.~H.}\ \bibnamefont
  {von Gr\"unberg}}, \bibinfo {author} {\bibfnamefont {P.}~\bibnamefont
  {Keim}}, \bibinfo {author} {\bibfnamefont {K.}~\bibnamefont {Zahn}}, \ and\
  \bibinfo {author} {\bibfnamefont {G.}~\bibnamefont {Maret}},\ }\bibfield
  {title} {\enquote {\bibinfo {title} {Elastic behavior of a two-dimensional
  crystal near melting},}\ }\href {\doibase 10.1103/PhysRevLett.93.255703}
  {\bibfield  {journal} {\bibinfo  {journal} {Phys. Rev. Lett.}\ }\textbf
  {\bibinfo {volume} {93}},\ \bibinfo {pages} {255703} (\bibinfo {year}
  {2004})}\BibitemShut {NoStop}%
\bibitem [{\citenamefont {Marcus}\ and\ \citenamefont {Rice}(1996)}]{Marcus96}%
  \BibitemOpen
  \bibfield  {author} {\bibinfo {author} {\bibfnamefont {A.~H.}\ \bibnamefont
  {Marcus}}\ and\ \bibinfo {author} {\bibfnamefont {S.~A.}\ \bibnamefont
  {Rice}},\ }\bibfield  {title} {\enquote {\bibinfo {title} {Observations of
  first-order liquid-to-hexatic and hexatic-to-solid phase transitions in a
  confined colloid suspension},}\ }\href {\doibase 10.1103/PhysRevLett.77.2577}
  {\bibfield  {journal} {\bibinfo  {journal} {Phys. Rev. Lett.}\ }\textbf
  {\bibinfo {volume} {77}},\ \bibinfo {pages} {2577--2580} (\bibinfo {year}
  {1996})}\BibitemShut {NoStop}%
\bibitem [{\citenamefont {Du}\ \emph {et~al.}(2017)\citenamefont {Du},
  \citenamefont {Doxastakis}, \citenamefont {Hilou},\ and\ \citenamefont
  {Biswal}}]{Du17}%
  \BibitemOpen
  \bibfield  {author} {\bibinfo {author} {\bibfnamefont {D.}~\bibnamefont
  {Du}}, \bibinfo {author} {\bibfnamefont {M.}~\bibnamefont {Doxastakis}},
  \bibinfo {author} {\bibfnamefont {E.}~\bibnamefont {Hilou}}, \ and\ \bibinfo
  {author} {\bibfnamefont {S.~L.}\ \bibnamefont {Biswal}},\ }\bibfield  {title}
  {\enquote {\bibinfo {title} {Two-dimensional melting of colloids with
  long-range attractive interactions},}\ }\href {\doibase 10.1039/C6SM02131J}
  {\bibfield  {journal} {\bibinfo  {journal} {Soft Matter}\ }\textbf {\bibinfo
  {volume} {13}},\ \bibinfo {pages} {1548--1553} (\bibinfo {year}
  {2017})}\BibitemShut {NoStop}%
\bibitem [{\citenamefont {Cha}\ and\ \citenamefont {Fertig}(1995)}]{Cha95}%
  \BibitemOpen
  \bibfield  {author} {\bibinfo {author} {\bibfnamefont {M.-C.}\ \bibnamefont
  {Cha}}\ and\ \bibinfo {author} {\bibfnamefont {H.~A.}\ \bibnamefont
  {Fertig}},\ }\bibfield  {title} {\enquote {\bibinfo {title} {Disorder-induced
  phase transitions in two-dimensional crystals},}\ }\href {\doibase
  10.1103/PhysRevLett.74.4867} {\bibfield  {journal} {\bibinfo  {journal}
  {Phys. Rev. Lett.}\ }\textbf {\bibinfo {volume} {74}},\ \bibinfo {pages}
  {4867--4870} (\bibinfo {year} {1995})}\BibitemShut {NoStop}%
\bibitem [{\citenamefont {Reichhardt}\ and\ \citenamefont
  {Reichhardt}(2017)}]{Reichhardt17}%
  \BibitemOpen
  \bibfield  {author} {\bibinfo {author} {\bibfnamefont {C.}~\bibnamefont
  {Reichhardt}}\ and\ \bibinfo {author} {\bibfnamefont {C.~J.~Olson}\
  \bibnamefont {Reichhardt}},\ }\bibfield  {title} {\enquote {\bibinfo {title}
  {Depinning and nonequilibrium dynamic phases of particle assemblies driven
  over random and ordered substrates: a review},}\ }\href {\doibase
  10.1088/1361-6633/80/2/026501} {\bibfield  {journal} {\bibinfo  {journal}
  {Rep. Prog. Phys.}\ }\textbf {\bibinfo {volume} {80}},\ \bibinfo {pages}
  {026501} (\bibinfo {year} {2017})}\BibitemShut {NoStop}%
\bibitem [{\citenamefont {Nelson}(1983)}]{Nelson83}%
  \BibitemOpen
  \bibfield  {author} {\bibinfo {author} {\bibfnamefont {D.~R.}\ \bibnamefont
  {Nelson}},\ }\bibfield  {title} {\enquote {\bibinfo {title} {Reentrant
  melting in solid films with quenched random impurities},}\ }\href {\doibase
  10.1103/PhysRevB.27.2902} {\bibfield  {journal} {\bibinfo  {journal} {Phys.
  Rev. B}\ }\textbf {\bibinfo {volume} {27}},\ \bibinfo {pages} {2902--2914}
  (\bibinfo {year} {1983})}\BibitemShut {NoStop}%
\bibitem [{\citenamefont {Deutschl\"ander}\ \emph {et~al.}(2013)\citenamefont
  {Deutschl\"ander}, \citenamefont {Horn}, \citenamefont {L\"owen},
  \citenamefont {Maret},\ and\ \citenamefont {Keim}}]{Deutschlander13}%
  \BibitemOpen
  \bibfield  {author} {\bibinfo {author} {\bibfnamefont {S.}~\bibnamefont
  {Deutschl\"ander}}, \bibinfo {author} {\bibfnamefont {T.}~\bibnamefont
  {Horn}}, \bibinfo {author} {\bibfnamefont {H.}~\bibnamefont {L\"owen}},
  \bibinfo {author} {\bibfnamefont {G.}~\bibnamefont {Maret}}, \ and\ \bibinfo
  {author} {\bibfnamefont {P.}~\bibnamefont {Keim}},\ }\bibfield  {title}
  {\enquote {\bibinfo {title} {Two-dimensional melting under quenched
  disorder},}\ }\href {\doibase 10.1103/PhysRevLett.111.098301} {\bibfield
  {journal} {\bibinfo  {journal} {Phys. Rev. Lett.}\ }\textbf {\bibinfo
  {volume} {111}},\ \bibinfo {pages} {098301} (\bibinfo {year}
  {2013})}\BibitemShut {NoStop}%
\bibitem [{\citenamefont {Reichhardt}\ and\ \citenamefont
  {Olson}(2002)}]{Reichhardt02}%
  \BibitemOpen
  \bibfield  {author} {\bibinfo {author} {\bibfnamefont {C.}~\bibnamefont
  {Reichhardt}}\ and\ \bibinfo {author} {\bibfnamefont {C.~J.}\ \bibnamefont
  {Olson}},\ }\bibfield  {title} {\enquote {\bibinfo {title} {Colloidal
  dynamics on disordered substrates},}\ }\href {\doibase
  10.1103/PhysRevLett.89.078301} {\bibfield  {journal} {\bibinfo  {journal}
  {Phys. Rev. Lett.}\ }\textbf {\bibinfo {volume} {89}},\ \bibinfo {pages}
  {078301} (\bibinfo {year} {2002})}\BibitemShut {NoStop}%
\bibitem [{\citenamefont {Pertsinidis}\ and\ \citenamefont
  {Ling}(2008)}]{Pertsinidis08}%
  \BibitemOpen
  \bibfield  {author} {\bibinfo {author} {\bibfnamefont {A.}~\bibnamefont
  {Pertsinidis}}\ and\ \bibinfo {author} {\bibfnamefont {X.~S.}\ \bibnamefont
  {Ling}},\ }\bibfield  {title} {\enquote {\bibinfo {title} {Statics and
  dynamics of {2D} colloidal crystals in a random pinning potential},}\ }\href
  {\doibase 10.1103/PhysRevLett.100.028303} {\bibfield  {journal} {\bibinfo
  {journal} {Phys. Rev. Lett.}\ }\textbf {\bibinfo {volume} {100}},\ \bibinfo
  {pages} {028303} (\bibinfo {year} {2008})}\BibitemShut {NoStop}%
\bibitem [{\citenamefont {Fily}\ \emph {et~al.}(2010)\citenamefont {Fily},
  \citenamefont {Olive}, \citenamefont {Di~Scala},\ and\ \citenamefont
  {Soret}}]{Fily10}%
  \BibitemOpen
  \bibfield  {author} {\bibinfo {author} {\bibfnamefont {Y.}~\bibnamefont
  {Fily}}, \bibinfo {author} {\bibfnamefont {E.}~\bibnamefont {Olive}},
  \bibinfo {author} {\bibfnamefont {N.}~\bibnamefont {Di~Scala}}, \ and\
  \bibinfo {author} {\bibfnamefont {J.~C.}\ \bibnamefont {Soret}},\ }\bibfield
  {title} {\enquote {\bibinfo {title} {Critical behavior of plastic depinning
  of vortex lattices in two dimensions: Molecular dynamics simulations},}\
  }\href {\doibase 10.1103/PhysRevB.82.134519} {\bibfield  {journal} {\bibinfo
  {journal} {Phys. Rev. B}\ }\textbf {\bibinfo {volume} {82}},\ \bibinfo
  {pages} {134519} (\bibinfo {year} {2010})}\BibitemShut {NoStop}%
\bibitem [{\citenamefont {Di~Scala}\ \emph {et~al.}(2012)\citenamefont
  {Di~Scala}, \citenamefont {Olive}, \citenamefont {Lansac}, \citenamefont
  {Fily},\ and\ \citenamefont {Soret}}]{DiScala12}%
  \BibitemOpen
  \bibfield  {author} {\bibinfo {author} {\bibfnamefont {N.}~\bibnamefont
  {Di~Scala}}, \bibinfo {author} {\bibfnamefont {E.}~\bibnamefont {Olive}},
  \bibinfo {author} {\bibfnamefont {Y.}~\bibnamefont {Lansac}}, \bibinfo
  {author} {\bibfnamefont {Y.}~\bibnamefont {Fily}}, \ and\ \bibinfo {author}
  {\bibfnamefont {J.~C.}\ \bibnamefont {Soret}},\ }\bibfield  {title} {\enquote
  {\bibinfo {title} {The elastic depinning transition of vortex lattices in two
  dimensions},}\ }\href {\doibase 10.1088/1367-2630/14/12/123027} {\bibfield
  {journal} {\bibinfo  {journal} {New J. Phys.}\ }\textbf {\bibinfo {volume}
  {14}},\ \bibinfo {pages} {123027} (\bibinfo {year} {2012})}\BibitemShut
  {NoStop}%
\bibitem [{\citenamefont {Reichhardt}\ \emph {et~al.}(2015)\citenamefont
  {Reichhardt}, \citenamefont {Ray},\ and\ \citenamefont
  {Reichhardt}}]{Reichhardt15}%
  \BibitemOpen
  \bibfield  {author} {\bibinfo {author} {\bibfnamefont {C.}~\bibnamefont
  {Reichhardt}}, \bibinfo {author} {\bibfnamefont {D.}~\bibnamefont {Ray}}, \
  and\ \bibinfo {author} {\bibfnamefont {C.~J.~Olson}\ \bibnamefont
  {Reichhardt}},\ }\bibfield  {title} {\enquote {\bibinfo {title} {Collective
  transport properties of driven skyrmions with random disorder},}\ }\href
  {\doibase 10.1103/PhysRevLett.114.217202} {\bibfield  {journal} {\bibinfo
  {journal} {Phys. Rev. Lett.}\ }\textbf {\bibinfo {volume} {114}},\ \bibinfo
  {pages} {217202} (\bibinfo {year} {2015})}\BibitemShut {NoStop}%
\bibitem [{\citenamefont {Pippard}(1969)}]{Pippard69}%
  \BibitemOpen
  \bibfield  {author} {\bibinfo {author} {\bibfnamefont {A.~B.}\ \bibnamefont
  {Pippard}},\ }\bibfield  {title} {\enquote {\bibinfo {title} {A possible
  mechanism for peak effect in type 2 superconductors},}\ }\href {\doibase
  10.1080/14786436908217779} {\bibfield  {journal} {\bibinfo  {journal} {Phil.
  Mag.}\ }\textbf {\bibinfo {volume} {19}},\ \bibinfo {pages} {217} (\bibinfo
  {year} {1969})}\BibitemShut {NoStop}%
\bibitem [{\citenamefont {Kes}\ and\ \citenamefont {Tsuei}(1983)}]{Kes83}%
  \BibitemOpen
  \bibfield  {author} {\bibinfo {author} {\bibfnamefont {P.~H.}\ \bibnamefont
  {Kes}}\ and\ \bibinfo {author} {\bibfnamefont {C.~C.}\ \bibnamefont
  {Tsuei}},\ }\bibfield  {title} {\enquote {\bibinfo {title} {Two-dimensional
  collective flux pinning, defects, and structural relaxation in amorphous
  superconducting films},}\ }\href {\doibase 10.1103/PhysRevB.28.5126}
  {\bibfield  {journal} {\bibinfo  {journal} {Phys. Rev. B}\ }\textbf {\bibinfo
  {volume} {28}},\ \bibinfo {pages} {5126--5139} (\bibinfo {year}
  {1983})}\BibitemShut {NoStop}%
\bibitem [{\citenamefont {W\"ordenweber}\ \emph {et~al.}(1986)\citenamefont
  {W\"ordenweber}, \citenamefont {Kes},\ and\ \citenamefont
  {Tsuei}}]{Wordenweber86a}%
  \BibitemOpen
  \bibfield  {author} {\bibinfo {author} {\bibfnamefont {R.}~\bibnamefont
  {W\"ordenweber}}, \bibinfo {author} {\bibfnamefont {P.~H.}\ \bibnamefont
  {Kes}}, \ and\ \bibinfo {author} {\bibfnamefont {C.~C.}\ \bibnamefont
  {Tsuei}},\ }\bibfield  {title} {\enquote {\bibinfo {title} {Peak and history
  effects in two-dimensional collective flux pinning},}\ }\href {\doibase
  10.1103/PhysRevB.33.3172} {\bibfield  {journal} {\bibinfo  {journal} {Phys.
  Rev. B}\ }\textbf {\bibinfo {volume} {33}},\ \bibinfo {pages} {3172--3180}
  (\bibinfo {year} {1986})}\BibitemShut {NoStop}%
\bibitem [{\citenamefont {Bhattacharya}\ and\ \citenamefont
  {Higgins}(1993)}]{Bhattacharya93}%
  \BibitemOpen
  \bibfield  {author} {\bibinfo {author} {\bibfnamefont {S.}~\bibnamefont
  {Bhattacharya}}\ and\ \bibinfo {author} {\bibfnamefont {M.~J.}\ \bibnamefont
  {Higgins}},\ }\bibfield  {title} {\enquote {\bibinfo {title} {Dynamics of a
  disordered flux line lattice},}\ }\href {\doibase
  10.1103/PhysRevLett.70.2617} {\bibfield  {journal} {\bibinfo  {journal}
  {Phys. Rev. Lett.}\ }\textbf {\bibinfo {volume} {70}},\ \bibinfo {pages}
  {2617--2620} (\bibinfo {year} {1993})}\BibitemShut {NoStop}%
\bibitem [{\citenamefont {Kwok}\ \emph {et~al.}(1994)\citenamefont {Kwok},
  \citenamefont {Fendrich}, \citenamefont {van~der Beek},\ and\ \citenamefont
  {Crabtree}}]{Kwok94}%
  \BibitemOpen
  \bibfield  {author} {\bibinfo {author} {\bibfnamefont {W.~K.}\ \bibnamefont
  {Kwok}}, \bibinfo {author} {\bibfnamefont {J.~A.}\ \bibnamefont {Fendrich}},
  \bibinfo {author} {\bibfnamefont {C.~J.}\ \bibnamefont {van~der Beek}}, \
  and\ \bibinfo {author} {\bibfnamefont {G.~W.}\ \bibnamefont {Crabtree}},\
  }\bibfield  {title} {\enquote {\bibinfo {title} {Peak effect as a precursor
  to vortex lattice melting in single crystal {YBa$_2$Cu$_3$O$_{7-\delta}$}},}\
  }\href {\doibase 10.1103/PhysRevLett.73.2614} {\bibfield  {journal} {\bibinfo
   {journal} {Phys. Rev. Lett.}\ }\textbf {\bibinfo {volume} {73}},\ \bibinfo
  {pages} {2614--2617} (\bibinfo {year} {1994})}\BibitemShut {NoStop}%
\bibitem [{\citenamefont {Gammel}\ \emph {et~al.}(1998)\citenamefont {Gammel},
  \citenamefont {Yaron}, \citenamefont {Ramirez}, \citenamefont {Bishop},
  \citenamefont {Chang}, \citenamefont {Ruel}, \citenamefont {Pfeiffer},
  \citenamefont {Bucher}, \citenamefont {D'Anna}, \citenamefont {Huse},
  \citenamefont {Mortensen}, \citenamefont {Eskildsen},\ and\ \citenamefont
  {Kes}}]{Gammel98}%
  \BibitemOpen
  \bibfield  {author} {\bibinfo {author} {\bibfnamefont {P.~L.}\ \bibnamefont
  {Gammel}}, \bibinfo {author} {\bibfnamefont {U.}~\bibnamefont {Yaron}},
  \bibinfo {author} {\bibfnamefont {A.~P.}\ \bibnamefont {Ramirez}}, \bibinfo
  {author} {\bibfnamefont {D.~J.}\ \bibnamefont {Bishop}}, \bibinfo {author}
  {\bibfnamefont {A.~M.}\ \bibnamefont {Chang}}, \bibinfo {author}
  {\bibfnamefont {R.}~\bibnamefont {Ruel}}, \bibinfo {author} {\bibfnamefont
  {L.~N.}\ \bibnamefont {Pfeiffer}}, \bibinfo {author} {\bibfnamefont
  {E.}~\bibnamefont {Bucher}}, \bibinfo {author} {\bibfnamefont
  {G.}~\bibnamefont {D'Anna}}, \bibinfo {author} {\bibfnamefont {D.~A.}\
  \bibnamefont {Huse}}, \bibinfo {author} {\bibfnamefont {K.}~\bibnamefont
  {Mortensen}}, \bibinfo {author} {\bibfnamefont {M.~R.}\ \bibnamefont
  {Eskildsen}}, \ and\ \bibinfo {author} {\bibfnamefont {P.~H.}\ \bibnamefont
  {Kes}},\ }\bibfield  {title} {\enquote {\bibinfo {title} {Structure and
  correlations of the flux line lattice in crystalline {Nb} through the peak
  effect},}\ }\href {\doibase 10.1103/PhysRevLett.80.833} {\bibfield  {journal}
  {\bibinfo  {journal} {Phys. Rev. Lett.}\ }\textbf {\bibinfo {volume} {80}},\
  \bibinfo {pages} {833--836} (\bibinfo {year} {1998})}\BibitemShut {NoStop}%
\bibitem [{\citenamefont {Banerjee}\ \emph {et~al.}(1999)\citenamefont
  {Banerjee}, \citenamefont {Patil}, \citenamefont {Ramakrishnan},
  \citenamefont {Grover}, \citenamefont {Bhattacharya}, \citenamefont {Mishra},
  \citenamefont {Ravikumar}, \citenamefont {Chandrasekhar~Rao}, \citenamefont
  {Sahni}, \citenamefont {Higgins}, \citenamefont {Tomy}, \citenamefont
  {Balakrishnan},\ and\ \citenamefont {Mck.~Paul}}]{Banerjee99}%
  \BibitemOpen
  \bibfield  {author} {\bibinfo {author} {\bibfnamefont {S.~S.}\ \bibnamefont
  {Banerjee}}, \bibinfo {author} {\bibfnamefont {N.~G.}\ \bibnamefont {Patil}},
  \bibinfo {author} {\bibfnamefont {S.}~\bibnamefont {Ramakrishnan}}, \bibinfo
  {author} {\bibfnamefont {A.~K.}\ \bibnamefont {Grover}}, \bibinfo {author}
  {\bibfnamefont {S.}~\bibnamefont {Bhattacharya}}, \bibinfo {author}
  {\bibfnamefont {P.~K.}\ \bibnamefont {Mishra}}, \bibinfo {author}
  {\bibfnamefont {G.}~\bibnamefont {Ravikumar}}, \bibinfo {author}
  {\bibfnamefont {T.~V.}\ \bibnamefont {Chandrasekhar~Rao}}, \bibinfo {author}
  {\bibfnamefont {V.~C.}\ \bibnamefont {Sahni}}, \bibinfo {author}
  {\bibfnamefont {M.~J.}\ \bibnamefont {Higgins}}, \bibinfo {author}
  {\bibfnamefont {C.~V.}\ \bibnamefont {Tomy}}, \bibinfo {author}
  {\bibfnamefont {G.}~\bibnamefont {Balakrishnan}}, \ and\ \bibinfo {author}
  {\bibfnamefont {D.}~\bibnamefont {Mck.~Paul}},\ }\bibfield  {title} {\enquote
  {\bibinfo {title} {Disorder, metastability, and history dependence in
  transformations of a vortex lattice},}\ }\href {\doibase
  10.1103/PhysRevB.59.6043} {\bibfield  {journal} {\bibinfo  {journal} {Phys.
  Rev. B}\ }\textbf {\bibinfo {volume} {59}},\ \bibinfo {pages} {6043--6046}
  (\bibinfo {year} {1999})}\BibitemShut {NoStop}%
\bibitem [{\citenamefont {Paltiel}\ \emph {et~al.}(2000)\citenamefont
  {Paltiel}, \citenamefont {Zeldov}, \citenamefont {Myasoedov}, \citenamefont
  {Shtrikman}, \citenamefont {Bhattacharya}, \citenamefont {Higgins},
  \citenamefont {Xiao}, \citenamefont {Andrei}, \citenamefont {Gammel},\ and\
  \citenamefont {Bishop}}]{Paltiel00a}%
  \BibitemOpen
  \bibfield  {author} {\bibinfo {author} {\bibfnamefont {Y.}~\bibnamefont
  {Paltiel}}, \bibinfo {author} {\bibfnamefont {E.}~\bibnamefont {Zeldov}},
  \bibinfo {author} {\bibfnamefont {Y.~N.}\ \bibnamefont {Myasoedov}}, \bibinfo
  {author} {\bibfnamefont {H.}~\bibnamefont {Shtrikman}}, \bibinfo {author}
  {\bibfnamefont {S.}~\bibnamefont {Bhattacharya}}, \bibinfo {author}
  {\bibfnamefont {M.~J.}\ \bibnamefont {Higgins}}, \bibinfo {author}
  {\bibfnamefont {Z.~L.}\ \bibnamefont {Xiao}}, \bibinfo {author}
  {\bibfnamefont {E.~Y.}\ \bibnamefont {Andrei}}, \bibinfo {author}
  {\bibfnamefont {P.~L.}\ \bibnamefont {Gammel}}, \ and\ \bibinfo {author}
  {\bibfnamefont {D.~J.}\ \bibnamefont {Bishop}},\ }\bibfield  {title}
  {\enquote {\bibinfo {title} {Dynamic instabilities and memory effects in
  vortex matter},}\ }\href {\doibase 10.1038/35000145} {\bibfield  {journal}
  {\bibinfo  {journal} {Nature (London)}\ }\textbf {\bibinfo {volume} {403}},\
  \bibinfo {pages} {398--401} (\bibinfo {year} {2000})}\BibitemShut {NoStop}%
\bibitem [{\citenamefont {Ling}\ \emph {et~al.}(2001)\citenamefont {Ling},
  \citenamefont {Park}, \citenamefont {McClain}, \citenamefont {Choi},
  \citenamefont {Dender},\ and\ \citenamefont {Lynn}}]{Ling01}%
  \BibitemOpen
  \bibfield  {author} {\bibinfo {author} {\bibfnamefont {X.~S.}\ \bibnamefont
  {Ling}}, \bibinfo {author} {\bibfnamefont {S.~R.}\ \bibnamefont {Park}},
  \bibinfo {author} {\bibfnamefont {B.~A.}\ \bibnamefont {McClain}}, \bibinfo
  {author} {\bibfnamefont {S.~M.}\ \bibnamefont {Choi}}, \bibinfo {author}
  {\bibfnamefont {D.~C.}\ \bibnamefont {Dender}}, \ and\ \bibinfo {author}
  {\bibfnamefont {J.~W.}\ \bibnamefont {Lynn}},\ }\bibfield  {title} {\enquote
  {\bibinfo {title} {Superheating and supercooling of vortex matter in a {Nb}
  single crystal: Direct evidence for a phase transition at the peak effect
  from neutron diffraction},}\ }\href {\doibase 10.1103/PhysRevLett.86.712}
  {\bibfield  {journal} {\bibinfo  {journal} {Phys. Rev. Lett.}\ }\textbf
  {\bibinfo {volume} {86}},\ \bibinfo {pages} {712--715} (\bibinfo {year}
  {2001})}\BibitemShut {NoStop}%
\bibitem [{\citenamefont {Troyanovski}\ \emph {et~al.}(2002)\citenamefont
  {Troyanovski}, \citenamefont {van Hecke}, \citenamefont {Saha}, \citenamefont
  {Aarts},\ and\ \citenamefont {Kes}}]{Troyanovski02}%
  \BibitemOpen
  \bibfield  {author} {\bibinfo {author} {\bibfnamefont {A.~M.}\ \bibnamefont
  {Troyanovski}}, \bibinfo {author} {\bibfnamefont {M.}~\bibnamefont {van
  Hecke}}, \bibinfo {author} {\bibfnamefont {N.}~\bibnamefont {Saha}}, \bibinfo
  {author} {\bibfnamefont {J.}~\bibnamefont {Aarts}}, \ and\ \bibinfo {author}
  {\bibfnamefont {P.~H.}\ \bibnamefont {Kes}},\ }\bibfield  {title} {\enquote
  {\bibinfo {title} {{STM} imaging of flux line arrangements in the peak effect
  regime},}\ }\href {\doibase 10.1103/PhysRevLett.89.147006} {\bibfield
  {journal} {\bibinfo  {journal} {Phys. Rev. Lett.}\ }\textbf {\bibinfo
  {volume} {89}},\ \bibinfo {pages} {147006} (\bibinfo {year}
  {2002})}\BibitemShut {NoStop}%
\bibitem [{\citenamefont {Hilke}\ \emph {et~al.}(2003)\citenamefont {Hilke},
  \citenamefont {Reid}, \citenamefont {Gagnon},\ and\ \citenamefont
  {Altounian}}]{Hilke03}%
  \BibitemOpen
  \bibfield  {author} {\bibinfo {author} {\bibfnamefont {M.}~\bibnamefont
  {Hilke}}, \bibinfo {author} {\bibfnamefont {S.}~\bibnamefont {Reid}},
  \bibinfo {author} {\bibfnamefont {R.}~\bibnamefont {Gagnon}}, \ and\ \bibinfo
  {author} {\bibfnamefont {Z.}~\bibnamefont {Altounian}},\ }\bibfield  {title}
  {\enquote {\bibinfo {title} {Peak effect and the phase diagram of moving
  vortices in {Fe}$_x${Ni}$_{1-x}${Zr}$_{2}$ superconducting glasses},}\ }\href
  {\doibase 10.1103/PhysRevLett.91.127004} {\bibfield  {journal} {\bibinfo
  {journal} {Phys. Rev. Lett.}\ }\textbf {\bibinfo {volume} {91}},\ \bibinfo
  {pages} {127004} (\bibinfo {year} {2003})}\BibitemShut {NoStop}%
\bibitem [{\citenamefont {Pasquini}\ \emph {et~al.}(2008)\citenamefont
  {Pasquini}, \citenamefont {Daroca}, \citenamefont {Chiliotte}, \citenamefont
  {Lozano},\ and\ \citenamefont {Bekeris}}]{Pasquini08}%
  \BibitemOpen
  \bibfield  {author} {\bibinfo {author} {\bibfnamefont {G.}~\bibnamefont
  {Pasquini}}, \bibinfo {author} {\bibfnamefont {D.~P\'erez}\ \bibnamefont
  {Daroca}}, \bibinfo {author} {\bibfnamefont {C.}~\bibnamefont {Chiliotte}},
  \bibinfo {author} {\bibfnamefont {G.~S.}\ \bibnamefont {Lozano}}, \ and\
  \bibinfo {author} {\bibfnamefont {V.}~\bibnamefont {Bekeris}},\ }\bibfield
  {title} {\enquote {\bibinfo {title} {Ordered, disordered, and coexistent
  stable vortex lattices in {NbSe}$_2$ single crystals},}\ }\href {\doibase
  10.1103/PhysRevLett.100.247003} {\bibfield  {journal} {\bibinfo  {journal}
  {Phys. Rev. Lett.}\ }\textbf {\bibinfo {volume} {100}},\ \bibinfo {pages}
  {247003} (\bibinfo {year} {2008})}\BibitemShut {NoStop}%
\bibitem [{\citenamefont {Okuma}\ \emph {et~al.}(2008)\citenamefont {Okuma},
  \citenamefont {Kashiro}, \citenamefont {Suzuki},\ and\ \citenamefont
  {Kokubo}}]{Okuma08}%
  \BibitemOpen
  \bibfield  {author} {\bibinfo {author} {\bibfnamefont {S.}~\bibnamefont
  {Okuma}}, \bibinfo {author} {\bibfnamefont {K.}~\bibnamefont {Kashiro}},
  \bibinfo {author} {\bibfnamefont {Y.}~\bibnamefont {Suzuki}}, \ and\ \bibinfo
  {author} {\bibfnamefont {N.}~\bibnamefont {Kokubo}},\ }\bibfield  {title}
  {\enquote {\bibinfo {title} {Order-disorder transition of vortex matter in
  {a-Mo$_x$Ge$_{1-x}$} films probed by noise},}\ }\href {\doibase
  10.1103/PhysRevB.77.212505} {\bibfield  {journal} {\bibinfo  {journal} {Phys.
  Rev. B}\ }\textbf {\bibinfo {volume} {77}},\ \bibinfo {pages} {212505}
  (\bibinfo {year} {2008})}\BibitemShut {NoStop}%
\bibitem [{\citenamefont {Wigner}(1934)}]{Wigner34}%
  \BibitemOpen
  \bibfield  {author} {\bibinfo {author} {\bibfnamefont {E.}~\bibnamefont
  {Wigner}},\ }\bibfield  {title} {\enquote {\bibinfo {title} {On the
  interaction of electrons in metals},}\ }\href {\doibase
  10.1103/PhysRev.46.1002} {\bibfield  {journal} {\bibinfo  {journal} {Phys.
  Rev.}\ }\textbf {\bibinfo {volume} {46}},\ \bibinfo {pages} {1002--1011}
  (\bibinfo {year} {1934})}\BibitemShut {NoStop}%
\bibitem [{\citenamefont {Andrei}\ \emph {et~al.}(1988)\citenamefont {Andrei},
  \citenamefont {Deville}, \citenamefont {Glattli}, \citenamefont {Williams},
  \citenamefont {Paris},\ and\ \citenamefont {Etienne}}]{Andrei88}%
  \BibitemOpen
  \bibfield  {author} {\bibinfo {author} {\bibfnamefont {E.~Y.}\ \bibnamefont
  {Andrei}}, \bibinfo {author} {\bibfnamefont {G.}~\bibnamefont {Deville}},
  \bibinfo {author} {\bibfnamefont {D.~C.}\ \bibnamefont {Glattli}}, \bibinfo
  {author} {\bibfnamefont {F.~I.~B.}\ \bibnamefont {Williams}}, \bibinfo
  {author} {\bibfnamefont {E.}~\bibnamefont {Paris}}, \ and\ \bibinfo {author}
  {\bibfnamefont {B.}~\bibnamefont {Etienne}},\ }\bibfield  {title} {\enquote
  {\bibinfo {title} {Observation of a magnetically induced {W}igner solid},}\
  }\href {\doibase 10.1103/PhysRevLett.60.2765} {\bibfield  {journal} {\bibinfo
   {journal} {Phys. Rev. Lett.}\ }\textbf {\bibinfo {volume} {60}},\ \bibinfo
  {pages} {2765--2768} (\bibinfo {year} {1988})}\BibitemShut {NoStop}%
\bibitem [{\citenamefont {Goldman}\ \emph {et~al.}(1990)\citenamefont
  {Goldman}, \citenamefont {Santos}, \citenamefont {Shayegan},\ and\
  \citenamefont {Cunningham}}]{Goldman90}%
  \BibitemOpen
  \bibfield  {author} {\bibinfo {author} {\bibfnamefont {V.~J.}\ \bibnamefont
  {Goldman}}, \bibinfo {author} {\bibfnamefont {M}~\bibnamefont {Santos}},
  \bibinfo {author} {\bibfnamefont {M}~\bibnamefont {Shayegan}}, \ and\
  \bibinfo {author} {\bibfnamefont {J.~E.}\ \bibnamefont {Cunningham}},\
  }\bibfield  {title} {\enquote {\bibinfo {title} {Evidence for two-dimentional
  quantum {W}igner crystal},}\ }\href {\doibase 10.1103/PhysRevLett.65.2189}
  {\bibfield  {journal} {\bibinfo  {journal} {Phys. Rev. Lett.}\ }\textbf
  {\bibinfo {volume} {65}},\ \bibinfo {pages} {2189--2192} (\bibinfo {year}
  {1990})}\BibitemShut {NoStop}%
\bibitem [{\citenamefont {Jiang}\ \emph {et~al.}(1991)\citenamefont {Jiang},
  \citenamefont {Stormer}, \citenamefont {Tsui}, \citenamefont {Pfeiffer},\
  and\ \citenamefont {West}}]{Jiang91}%
  \BibitemOpen
  \bibfield  {author} {\bibinfo {author} {\bibfnamefont {H.~W.}\ \bibnamefont
  {Jiang}}, \bibinfo {author} {\bibfnamefont {H.~L.}\ \bibnamefont {Stormer}},
  \bibinfo {author} {\bibfnamefont {D.~C.}\ \bibnamefont {Tsui}}, \bibinfo
  {author} {\bibfnamefont {L.~N.}\ \bibnamefont {Pfeiffer}}, \ and\ \bibinfo
  {author} {\bibfnamefont {K.~W.}\ \bibnamefont {West}},\ }\bibfield  {title}
  {\enquote {\bibinfo {title} {Magnetotransport studies of the insulating phase
  around \ensuremath{\nu}=1/5 {L}andau-level filling},}\ }\href {\doibase
  10.1103/PhysRevB.44.8107} {\bibfield  {journal} {\bibinfo  {journal} {Phys.
  Rev. B}\ }\textbf {\bibinfo {volume} {44}},\ \bibinfo {pages} {8107--8114}
  (\bibinfo {year} {1991})}\BibitemShut {NoStop}%
\bibitem [{\citenamefont {Williams}\ \emph {et~al.}(1991)\citenamefont
  {Williams}, \citenamefont {Wright}, \citenamefont {Clark}, \citenamefont
  {Andrei}, \citenamefont {Deville}, \citenamefont {Glattli}, \citenamefont
  {Probst}, \citenamefont {Etienne}, \citenamefont {Dorin}, \citenamefont
  {Foxon},\ and\ \citenamefont {Harris}}]{Williams91}%
  \BibitemOpen
  \bibfield  {author} {\bibinfo {author} {\bibfnamefont {F.~I.~B.}\
  \bibnamefont {Williams}}, \bibinfo {author} {\bibfnamefont {P.~A.}\
  \bibnamefont {Wright}}, \bibinfo {author} {\bibfnamefont {R.~G.}\
  \bibnamefont {Clark}}, \bibinfo {author} {\bibfnamefont {E.~Y.}\ \bibnamefont
  {Andrei}}, \bibinfo {author} {\bibfnamefont {G.}~\bibnamefont {Deville}},
  \bibinfo {author} {\bibfnamefont {D.~C.}\ \bibnamefont {Glattli}}, \bibinfo
  {author} {\bibfnamefont {O.}~\bibnamefont {Probst}}, \bibinfo {author}
  {\bibfnamefont {B.}~\bibnamefont {Etienne}}, \bibinfo {author} {\bibfnamefont
  {C.}~\bibnamefont {Dorin}}, \bibinfo {author} {\bibfnamefont {C.~T.}\
  \bibnamefont {Foxon}}, \ and\ \bibinfo {author} {\bibfnamefont {J.~J.}\
  \bibnamefont {Harris}},\ }\bibfield  {title} {\enquote {\bibinfo {title}
  {Conduction threshold and pinning frequency of magnetically induced {W}igner
  solid},}\ }\href {\doibase 10.1103/PhysRevLett.66.3285} {\bibfield  {journal}
  {\bibinfo  {journal} {Phys. Rev. Lett.}\ }\textbf {\bibinfo {volume} {66}},\
  \bibinfo {pages} {3285--3288} (\bibinfo {year} {1991})}\BibitemShut {NoStop}%
\bibitem [{\citenamefont {Piacente}\ and\ \citenamefont
  {Peeters}(2005)}]{Piacente05}%
  \BibitemOpen
  \bibfield  {author} {\bibinfo {author} {\bibfnamefont {G.}~\bibnamefont
  {Piacente}}\ and\ \bibinfo {author} {\bibfnamefont {F.~M.}\ \bibnamefont
  {Peeters}},\ }\bibfield  {title} {\enquote {\bibinfo {title} {Pinning and
  depinning of a classic quasi-one-dimensional {W}igner crystal in the presence
  of a constriction},}\ }\href {\doibase 10.1103/PhysRevB.72.205208} {\bibfield
   {journal} {\bibinfo  {journal} {Phys. Rev. B}\ }\textbf {\bibinfo {volume}
  {72}},\ \bibinfo {pages} {205208} (\bibinfo {year} {2005})}\BibitemShut
  {NoStop}%
\bibitem [{\citenamefont {Jang}\ \emph {et~al.}(2017)\citenamefont {Jang},
  \citenamefont {Hunt}, \citenamefont {Pfeiffer}, \citenamefont {West},\ and\
  \citenamefont {Ashoori}}]{Jang17}%
  \BibitemOpen
  \bibfield  {author} {\bibinfo {author} {\bibfnamefont {J.}~\bibnamefont
  {Jang}}, \bibinfo {author} {\bibfnamefont {B.~M.}\ \bibnamefont {Hunt}},
  \bibinfo {author} {\bibfnamefont {L.~N.}\ \bibnamefont {Pfeiffer}}, \bibinfo
  {author} {\bibfnamefont {K.~W.}\ \bibnamefont {West}}, \ and\ \bibinfo
  {author} {\bibfnamefont {R.~C.}\ \bibnamefont {Ashoori}},\ }\bibfield
  {title} {\enquote {\bibinfo {title} {Sharp tunneling resonance from the
  vibrations of an electronic {Wigner} crystal},}\ }\href {\doibase
  10.1038/nphys3979} {\bibfield  {journal} {\bibinfo  {journal} {Nature Phys.}\
  }\textbf {\bibinfo {volume} {13}},\ \bibinfo {pages} {340 -- 344} (\bibinfo
  {year} {2017})}\BibitemShut {NoStop}%
\bibitem [{\citenamefont {Brussarski}\ \emph {et~al.}(2018)\citenamefont
  {Brussarski}, \citenamefont {Li}, \citenamefont {Kravchenko}, \citenamefont
  {Shashkin},\ and\ \citenamefont {Sarachik}}]{Brussarski18}%
  \BibitemOpen
  \bibfield  {author} {\bibinfo {author} {\bibfnamefont {P.}~\bibnamefont
  {Brussarski}}, \bibinfo {author} {\bibfnamefont {S.}~\bibnamefont {Li}},
  \bibinfo {author} {\bibfnamefont {S.~V.}\ \bibnamefont {Kravchenko}},
  \bibinfo {author} {\bibfnamefont {A.~A.}\ \bibnamefont {Shashkin}}, \ and\
  \bibinfo {author} {\bibfnamefont {M.~P.}\ \bibnamefont {Sarachik}},\
  }\bibfield  {title} {\enquote {\bibinfo {title} {Transport evidence for a
  sliding two-dimensional quantum electron solid},}\ }\href {\doibase
  10.1038/s41467-018-06332-9} {\bibfield  {journal} {\bibinfo  {journal}
  {Nature Commun.}\ }\textbf {\bibinfo {volume} {9}},\ \bibinfo {pages} {3803}
  (\bibinfo {year} {2018})}\BibitemShut {NoStop}%
\bibitem [{\citenamefont {Hatke}\ \emph {et~al.}(2019)\citenamefont {Hatke},
  \citenamefont {Deng}, \citenamefont {Liu}, \citenamefont {Engel},
  \citenamefont {Pfeiffer}, \citenamefont {West}, \citenamefont {Baldwin},\
  and\ \citenamefont {Shayegan}}]{Hatke19}%
  \BibitemOpen
  \bibfield  {author} {\bibinfo {author} {\bibfnamefont {A.~T.}\ \bibnamefont
  {Hatke}}, \bibinfo {author} {\bibfnamefont {H.}~\bibnamefont {Deng}},
  \bibinfo {author} {\bibfnamefont {Y.}~\bibnamefont {Liu}}, \bibinfo {author}
  {\bibfnamefont {L.~W.}\ \bibnamefont {Engel}}, \bibinfo {author}
  {\bibfnamefont {L.~N.}\ \bibnamefont {Pfeiffer}}, \bibinfo {author}
  {\bibfnamefont {K.~W.}\ \bibnamefont {West}}, \bibinfo {author}
  {\bibfnamefont {K.~W.}\ \bibnamefont {Baldwin}}, \ and\ \bibinfo {author}
  {\bibfnamefont {M.}~\bibnamefont {Shayegan}},\ }\bibfield  {title} {\enquote
  {\bibinfo {title} {Wigner solid pinning modes tuned by fractional quantum
  {H}all states of a nearby layer},}\ }\href {\doibase 10.1126/sciadv.aao2848}
  {\bibfield  {journal} {\bibinfo  {journal} {Sci. Adv.}\ }\textbf {\bibinfo
  {volume} {5}},\ \bibinfo {pages} {eaao2848} (\bibinfo {year}
  {2019})}\BibitemShut {NoStop}%
\bibitem [{\citenamefont {Hossain}\ \emph {et~al.}(2022)\citenamefont
  {Hossain}, \citenamefont {Ma}, \citenamefont {Villegas-Rosales},
  \citenamefont {Chung}, \citenamefont {Pfeiffer}, \citenamefont {West},
  \citenamefont {Baldwin},\ and\ \citenamefont {Shayegan}}]{Hossain22}%
  \BibitemOpen
  \bibfield  {author} {\bibinfo {author} {\bibfnamefont {Md.~S.}\ \bibnamefont
  {Hossain}}, \bibinfo {author} {\bibfnamefont {M.~K.}\ \bibnamefont {Ma}},
  \bibinfo {author} {\bibfnamefont {K.~A.}\ \bibnamefont {Villegas-Rosales}},
  \bibinfo {author} {\bibfnamefont {Y.~J.}\ \bibnamefont {Chung}}, \bibinfo
  {author} {\bibfnamefont {L.~N.}\ \bibnamefont {Pfeiffer}}, \bibinfo {author}
  {\bibfnamefont {K.~W.}\ \bibnamefont {West}}, \bibinfo {author}
  {\bibfnamefont {K.~W.}\ \bibnamefont {Baldwin}}, \ and\ \bibinfo {author}
  {\bibfnamefont {M.}~\bibnamefont {Shayegan}},\ }\bibfield  {title} {\enquote
  {\bibinfo {title} {Anisotropic two-dimensional disordered {W}igner solid},}\
  }\href {\doibase 10.1103/PhysRevLett.129.036601} {\bibfield  {journal}
  {\bibinfo  {journal} {Phys. Rev. Lett.}\ }\textbf {\bibinfo {volume} {129}},\
  \bibinfo {pages} {036601} (\bibinfo {year} {2022})}\BibitemShut {NoStop}%
\bibitem [{\citenamefont {Chen}\ \emph {et~al.}(2006)\citenamefont {Chen},
  \citenamefont {Sambandamurthy}, \citenamefont {Wang}, \citenamefont {Lewis},
  \citenamefont {Engel}, \citenamefont {Tsui}, \citenamefont {Ye},
  \citenamefont {Pfeiffer},\ and\ \citenamefont {West}}]{Chen06}%
  \BibitemOpen
  \bibfield  {author} {\bibinfo {author} {\bibfnamefont {Y.~P.}\ \bibnamefont
  {Chen}}, \bibinfo {author} {\bibfnamefont {G.}~\bibnamefont
  {Sambandamurthy}}, \bibinfo {author} {\bibfnamefont {Z.~H.}\ \bibnamefont
  {Wang}}, \bibinfo {author} {\bibfnamefont {R.~M.}\ \bibnamefont {Lewis}},
  \bibinfo {author} {\bibfnamefont {L.~W.}\ \bibnamefont {Engel}}, \bibinfo
  {author} {\bibfnamefont {D.~C.}\ \bibnamefont {Tsui}}, \bibinfo {author}
  {\bibfnamefont {P.~D.}\ \bibnamefont {Ye}}, \bibinfo {author} {\bibfnamefont
  {L.~N.}\ \bibnamefont {Pfeiffer}}, \ and\ \bibinfo {author} {\bibfnamefont
  {K.~W.}\ \bibnamefont {West}},\ }\bibfield  {title} {\enquote {\bibinfo
  {title} {Melting of a {2D} quantum electron solid in high magnetic field},}\
  }\href {\doibase 10.1038/nphys322} {\bibfield  {journal} {\bibinfo  {journal}
  {Nat. Phys.}\ }\textbf {\bibinfo {volume} {2}},\ \bibinfo {pages} {452--455}
  (\bibinfo {year} {2006})}\BibitemShut {NoStop}%
\bibitem [{\citenamefont {Knighton}\ \emph {et~al.}(2018)\citenamefont
  {Knighton}, \citenamefont {Wu}, \citenamefont {Huang}, \citenamefont
  {Serafin}, \citenamefont {Xia}, \citenamefont {Pfeiffer},\ and\ \citenamefont
  {West}}]{Knighton18}%
  \BibitemOpen
  \bibfield  {author} {\bibinfo {author} {\bibfnamefont {T.}~\bibnamefont
  {Knighton}}, \bibinfo {author} {\bibfnamefont {Z.}~\bibnamefont {Wu}},
  \bibinfo {author} {\bibfnamefont {J.}~\bibnamefont {Huang}}, \bibinfo
  {author} {\bibfnamefont {A.}~\bibnamefont {Serafin}}, \bibinfo {author}
  {\bibfnamefont {J.~S.}\ \bibnamefont {Xia}}, \bibinfo {author} {\bibfnamefont
  {L.~N.}\ \bibnamefont {Pfeiffer}}, \ and\ \bibinfo {author} {\bibfnamefont
  {K.~W.}\ \bibnamefont {West}},\ }\bibfield  {title} {\enquote {\bibinfo
  {title} {Evidence of two-stage melting of {Wigner} solids},}\ }\href
  {\doibase 10.1103/PhysRevB.97.085135} {\bibfield  {journal} {\bibinfo
  {journal} {Phys. Rev. B}\ }\textbf {\bibinfo {volume} {97}},\ \bibinfo
  {pages} {085135} (\bibinfo {year} {2018})}\BibitemShut {NoStop}%
\bibitem [{\citenamefont {Deng}\ \emph {et~al.}(2019)\citenamefont {Deng},
  \citenamefont {Pfeiffer}, \citenamefont {West}, \citenamefont {Baldwin},
  \citenamefont {Engel},\ and\ \citenamefont {Shayegan}}]{Deng19}%
  \BibitemOpen
  \bibfield  {author} {\bibinfo {author} {\bibfnamefont {H.}~\bibnamefont
  {Deng}}, \bibinfo {author} {\bibfnamefont {L.~N.}\ \bibnamefont {Pfeiffer}},
  \bibinfo {author} {\bibfnamefont {K.~W.}\ \bibnamefont {West}}, \bibinfo
  {author} {\bibfnamefont {K.~W.}\ \bibnamefont {Baldwin}}, \bibinfo {author}
  {\bibfnamefont {L.~W.}\ \bibnamefont {Engel}}, \ and\ \bibinfo {author}
  {\bibfnamefont {M.}~\bibnamefont {Shayegan}},\ }\bibfield  {title} {\enquote
  {\bibinfo {title} {Probing the melting of a two-dimensional quantum {W}igner
  crystal via its screening efficiency},}\ }\href {\doibase
  10.1103/PhysRevLett.122.116601} {\bibfield  {journal} {\bibinfo  {journal}
  {Phys. Rev. Lett.}\ }\textbf {\bibinfo {volume} {122}},\ \bibinfo {pages}
  {116601} (\bibinfo {year} {2019})}\BibitemShut {NoStop}%
\bibitem [{\citenamefont {Ma}\ \emph {et~al.}(2020)\citenamefont {Ma},
  \citenamefont {Villegas~Rosales}, \citenamefont {Deng}, \citenamefont
  {Chung}, \citenamefont {Pfeiffer}, \citenamefont {West}, \citenamefont
  {Baldwin}, \citenamefont {Winkler},\ and\ \citenamefont {Shayegan}}]{Ma20}%
  \BibitemOpen
  \bibfield  {author} {\bibinfo {author} {\bibfnamefont {M.~K.}\ \bibnamefont
  {Ma}}, \bibinfo {author} {\bibfnamefont {K.~A.}\ \bibnamefont
  {Villegas~Rosales}}, \bibinfo {author} {\bibfnamefont {H.}~\bibnamefont
  {Deng}}, \bibinfo {author} {\bibfnamefont {Y.~J.}\ \bibnamefont {Chung}},
  \bibinfo {author} {\bibfnamefont {L.~N.}\ \bibnamefont {Pfeiffer}}, \bibinfo
  {author} {\bibfnamefont {K.~W.}\ \bibnamefont {West}}, \bibinfo {author}
  {\bibfnamefont {K.~W.}\ \bibnamefont {Baldwin}}, \bibinfo {author}
  {\bibfnamefont {R.}~\bibnamefont {Winkler}}, \ and\ \bibinfo {author}
  {\bibfnamefont {M.}~\bibnamefont {Shayegan}},\ }\bibfield  {title} {\enquote
  {\bibinfo {title} {Thermal and quantum melting phase diagrams for a
  magnetic-field-induced {W}igner solid},}\ }\href {\doibase
  10.1103/PhysRevLett.125.036601} {\bibfield  {journal} {\bibinfo  {journal}
  {Phys. Rev. Lett.}\ }\textbf {\bibinfo {volume} {125}},\ \bibinfo {pages}
  {036601} (\bibinfo {year} {2020})}\BibitemShut {NoStop}%
\bibitem [{\citenamefont {Kim}\ \emph {et~al.}(2022)\citenamefont {Kim},
  \citenamefont {Bang}, \citenamefont {Lim}, \citenamefont {Lee}, \citenamefont
  {Hyun}, \citenamefont {Lee}, \citenamefont {Lee}, \citenamefont {Denlinger},
  \citenamefont {Huh}, \citenamefont {Kim}, \citenamefont {Song}, \citenamefont
  {Seo}, \citenamefont {Thapa}, \citenamefont {Kim}, \citenamefont {Lee},
  \citenamefont {Kim},\ and\ \citenamefont {Kim}}]{Kim22}%
  \BibitemOpen
  \bibfield  {author} {\bibinfo {author} {\bibfnamefont {S.}~\bibnamefont
  {Kim}}, \bibinfo {author} {\bibfnamefont {J.}~\bibnamefont {Bang}}, \bibinfo
  {author} {\bibfnamefont {C.}~\bibnamefont {Lim}}, \bibinfo {author}
  {\bibfnamefont {S.~Y.}\ \bibnamefont {Lee}}, \bibinfo {author} {\bibfnamefont
  {J.}~\bibnamefont {Hyun}}, \bibinfo {author} {\bibfnamefont {G.}~\bibnamefont
  {Lee}}, \bibinfo {author} {\bibfnamefont {Y.}~\bibnamefont {Lee}}, \bibinfo
  {author} {\bibfnamefont {J.~D.}\ \bibnamefont {Denlinger}}, \bibinfo {author}
  {\bibfnamefont {S.}~\bibnamefont {Huh}}, \bibinfo {author} {\bibfnamefont
  {C.}~\bibnamefont {Kim}}, \bibinfo {author} {\bibfnamefont {S.~Y.}\
  \bibnamefont {Song}}, \bibinfo {author} {\bibfnamefont {J.}~\bibnamefont
  {Seo}}, \bibinfo {author} {\bibfnamefont {D.}~\bibnamefont {Thapa}}, \bibinfo
  {author} {\bibfnamefont {S.-G.}\ \bibnamefont {Kim}}, \bibinfo {author}
  {\bibfnamefont {Y.~H.}\ \bibnamefont {Lee}}, \bibinfo {author} {\bibfnamefont
  {Y.}~\bibnamefont {Kim}}, \ and\ \bibinfo {author} {\bibfnamefont {S.~W.}\
  \bibnamefont {Kim}},\ }\bibfield  {title} {\enquote {\bibinfo {title}
  {Quantum electron liquid and its possible phase transition},}\ }\href
  {\doibase 10.1038/s41563-022-01353-8} {\bibfield  {journal} {\bibinfo
  {journal} {Nature Mater.}\ }\textbf {\bibinfo {volume} {21}},\ \bibinfo
  {pages} {1269--1274} (\bibinfo {year} {2022})}\BibitemShut {NoStop}%
\bibitem [{\citenamefont {Smole{\' n}ski}\ \emph {et~al.}(2021)\citenamefont
  {Smole{\' n}ski}, \citenamefont {Dolgirev}, \citenamefont {Kuhlenkamp},
  \citenamefont {Popert}, \citenamefont {Shimazaki}, \citenamefont {Back},
  \citenamefont {Lu}, \citenamefont {Kroner}, \citenamefont {Watanabe},
  \citenamefont {Taniguchi}, \citenamefont {Esterlis}, \citenamefont {Demler},\
  and\ \citenamefont {Imamoglu}}]{Smolenski21}%
  \BibitemOpen
  \bibfield  {author} {\bibinfo {author} {\bibfnamefont {T.}~\bibnamefont
  {Smole{\' n}ski}}, \bibinfo {author} {\bibfnamefont {P.~E.}\ \bibnamefont
  {Dolgirev}}, \bibinfo {author} {\bibfnamefont {C.}~\bibnamefont
  {Kuhlenkamp}}, \bibinfo {author} {\bibfnamefont {A.}~\bibnamefont {Popert}},
  \bibinfo {author} {\bibfnamefont {Y.}~\bibnamefont {Shimazaki}}, \bibinfo
  {author} {\bibfnamefont {P.}~\bibnamefont {Back}}, \bibinfo {author}
  {\bibfnamefont {X.}~\bibnamefont {Lu}}, \bibinfo {author} {\bibfnamefont
  {M.}~\bibnamefont {Kroner}}, \bibinfo {author} {\bibfnamefont
  {K.}~\bibnamefont {Watanabe}}, \bibinfo {author} {\bibfnamefont
  {T.}~\bibnamefont {Taniguchi}}, \bibinfo {author} {\bibfnamefont
  {I.}~\bibnamefont {Esterlis}}, \bibinfo {author} {\bibfnamefont
  {E.}~\bibnamefont {Demler}}, \ and\ \bibinfo {author} {\bibfnamefont
  {A.}~\bibnamefont {Imamoglu}},\ }\bibfield  {title} {\enquote {\bibinfo
  {title} {Signatures of {W}igner crystal of electrons in a monolayer
  semiconductor},}\ }\href {\doibase 10.1038/s41586-021-03590-4} {\bibfield
  {journal} {\bibinfo  {journal} {Nature (London)}\ }\textbf {\bibinfo {volume}
  {595}},\ \bibinfo {pages} {53--57} (\bibinfo {year} {2021})}\BibitemShut
  {NoStop}%
\bibitem [{\citenamefont {Xu}\ \emph {et~al.}(2020)\citenamefont {Xu},
  \citenamefont {Liu}, \citenamefont {Rhodes}, \citenamefont {Watanabe},
  \citenamefont {Taniguchi}, \citenamefont {Hone}, \citenamefont {Elser},
  \citenamefont {Mak},\ and\ \citenamefont {Shan}}]{Xu20}%
  \BibitemOpen
  \bibfield  {author} {\bibinfo {author} {\bibfnamefont {Y.}~\bibnamefont
  {Xu}}, \bibinfo {author} {\bibfnamefont {S.}~\bibnamefont {Liu}}, \bibinfo
  {author} {\bibfnamefont {D.~A.}\ \bibnamefont {Rhodes}}, \bibinfo {author}
  {\bibfnamefont {K.}~\bibnamefont {Watanabe}}, \bibinfo {author}
  {\bibfnamefont {T.}~\bibnamefont {Taniguchi}}, \bibinfo {author}
  {\bibfnamefont {J.}~\bibnamefont {Hone}}, \bibinfo {author} {\bibfnamefont
  {V.}~\bibnamefont {Elser}}, \bibinfo {author} {\bibfnamefont {K.~F.}\
  \bibnamefont {Mak}}, \ and\ \bibinfo {author} {\bibfnamefont
  {J.}~\bibnamefont {Shan}},\ }\bibfield  {title} {\enquote {\bibinfo {title}
  {Correlated insulating states at fractional fillings of moir{\` e}
  superlattices},}\ }\href {\doibase 10.1038/s41586-020-2868-6} {\bibfield
  {journal} {\bibinfo  {journal} {Nature (London)}\ }\textbf {\bibinfo {volume}
  {587}},\ \bibinfo {pages} {214--218} (\bibinfo {year} {2020})}\BibitemShut
  {NoStop}%
\bibitem [{\citenamefont {Li}\ \emph {et~al.}(2021)\citenamefont {Li},
  \citenamefont {Li}, \citenamefont {Regan}, \citenamefont {Wang},
  \citenamefont {Zhao}, \citenamefont {Kahn}, \citenamefont {Yumigeta},
  \citenamefont {Blei}, \citenamefont {Taniguchi}, \citenamefont {Watanabe},
  \citenamefont {Tongay}, \citenamefont {Zettl}, \citenamefont {Crommie},\ and\
  \citenamefont {Wang}}]{Li21}%
  \BibitemOpen
  \bibfield  {author} {\bibinfo {author} {\bibfnamefont {H.}~\bibnamefont
  {Li}}, \bibinfo {author} {\bibfnamefont {S.}~\bibnamefont {Li}}, \bibinfo
  {author} {\bibfnamefont {E.~C.}\ \bibnamefont {Regan}}, \bibinfo {author}
  {\bibfnamefont {D.}~\bibnamefont {Wang}}, \bibinfo {author} {\bibfnamefont
  {W.}~\bibnamefont {Zhao}}, \bibinfo {author} {\bibfnamefont {S.}~\bibnamefont
  {Kahn}}, \bibinfo {author} {\bibfnamefont {K.}~\bibnamefont {Yumigeta}},
  \bibinfo {author} {\bibfnamefont {M.}~\bibnamefont {Blei}}, \bibinfo {author}
  {\bibfnamefont {T.}~\bibnamefont {Taniguchi}}, \bibinfo {author}
  {\bibfnamefont {K.}~\bibnamefont {Watanabe}}, \bibinfo {author}
  {\bibfnamefont {S.}~\bibnamefont {Tongay}}, \bibinfo {author} {\bibfnamefont
  {A.}~\bibnamefont {Zettl}}, \bibinfo {author} {\bibfnamefont {M.~F.}\
  \bibnamefont {Crommie}}, \ and\ \bibinfo {author} {\bibfnamefont
  {F.}~\bibnamefont {Wang}},\ }\bibfield  {title} {\enquote {\bibinfo {title}
  {Imaging two-dimensional generalized {W}igner crystals},}\ }\href {\doibase
  10.1038/s41586-021-03874-9} {\bibfield  {journal} {\bibinfo  {journal}
  {Nature (London)}\ }\textbf {\bibinfo {volume} {597}},\ \bibinfo {pages}
  {650--654} (\bibinfo {year} {2021})}\BibitemShut {NoStop}%
\bibitem [{\citenamefont {Zhou}\ \emph {et~al.}(2021)\citenamefont {Zhou},
  \citenamefont {Sung}, \citenamefont {Brutschea}, \citenamefont {Esterlis},
  \citenamefont {Wang}, \citenamefont {Scuri}, \citenamefont {Gelly},
  \citenamefont {Heo}, \citenamefont {Taniguchi}, \citenamefont {Watanabe},
  \citenamefont {Zar{\' a}nd}, \citenamefont {Lukin}, \citenamefont {Kim},
  \citenamefont {Demler},\ and\ \citenamefont {Park}}]{Zhou21}%
  \BibitemOpen
  \bibfield  {author} {\bibinfo {author} {\bibfnamefont {Y.}~\bibnamefont
  {Zhou}}, \bibinfo {author} {\bibfnamefont {J.}~\bibnamefont {Sung}}, \bibinfo
  {author} {\bibfnamefont {E.}~\bibnamefont {Brutschea}}, \bibinfo {author}
  {\bibfnamefont {I.}~\bibnamefont {Esterlis}}, \bibinfo {author}
  {\bibfnamefont {Y.}~\bibnamefont {Wang}}, \bibinfo {author} {\bibfnamefont
  {G.}~\bibnamefont {Scuri}}, \bibinfo {author} {\bibfnamefont {R.~J.}\
  \bibnamefont {Gelly}}, \bibinfo {author} {\bibfnamefont {H.}~\bibnamefont
  {Heo}}, \bibinfo {author} {\bibfnamefont {T.}~\bibnamefont {Taniguchi}},
  \bibinfo {author} {\bibfnamefont {K.}~\bibnamefont {Watanabe}}, \bibinfo
  {author} {\bibfnamefont {G.}~\bibnamefont {Zar{\' a}nd}}, \bibinfo {author}
  {\bibfnamefont {M.~D.}\ \bibnamefont {Lukin}}, \bibinfo {author}
  {\bibfnamefont {P.}~\bibnamefont {Kim}}, \bibinfo {author} {\bibfnamefont
  {E.}~\bibnamefont {Demler}}, \ and\ \bibinfo {author} {\bibfnamefont
  {H.}~\bibnamefont {Park}},\ }\bibfield  {title} {\enquote {\bibinfo {title}
  {Bilayer {W}igner crystals in a transition metal dichalcogenide
  heterostructure},}\ }\href {\doibase 10.1038/s41586-021-03560-w} {\bibfield
  {journal} {\bibinfo  {journal} {Nature (London)}\ }\textbf {\bibinfo {volume}
  {595}},\ \bibinfo {pages} {48--52} (\bibinfo {year} {2021})}\BibitemShut
  {NoStop}%
\bibitem [{\citenamefont {Falson}\ \emph {et~al.}(2022)\citenamefont {Falson},
  \citenamefont {Sodemann}, \citenamefont {Skinner}, \citenamefont {Tabrea},
  \citenamefont {Kozuka}, \citenamefont {Tsukazaki}, \citenamefont {Kawasaki},
  \citenamefont {von Klitzing},\ and\ \citenamefont {Smet}}]{Falson22}%
  \BibitemOpen
  \bibfield  {author} {\bibinfo {author} {\bibfnamefont {J.}~\bibnamefont
  {Falson}}, \bibinfo {author} {\bibfnamefont {I.}~\bibnamefont {Sodemann}},
  \bibinfo {author} {\bibfnamefont {B.}~\bibnamefont {Skinner}}, \bibinfo
  {author} {\bibfnamefont {D.}~\bibnamefont {Tabrea}}, \bibinfo {author}
  {\bibfnamefont {Y.}~\bibnamefont {Kozuka}}, \bibinfo {author} {\bibfnamefont
  {A.}~\bibnamefont {Tsukazaki}}, \bibinfo {author} {\bibfnamefont
  {M.}~\bibnamefont {Kawasaki}}, \bibinfo {author} {\bibfnamefont
  {K.}~\bibnamefont {von Klitzing}}, \ and\ \bibinfo {author} {\bibfnamefont
  {J.~H.}\ \bibnamefont {Smet}},\ }\bibfield  {title} {\enquote {\bibinfo
  {title} {Competing correlated states around the zero-field {W}igner
  crystallization transition of electrons in two dimensions},}\ }\href
  {\doibase 10.1038/s41563-021-01166-1} {\bibfield  {journal} {\bibinfo
  {journal} {Nature Mater.}\ }\textbf {\bibinfo {volume} {21}},\ \bibinfo
  {pages} {311--316} (\bibinfo {year} {2022})}\BibitemShut {NoStop}%
\bibitem [{\citenamefont {Reichhardt}\ \emph {et~al.}(2001)\citenamefont
  {Reichhardt}, \citenamefont {Olson}, \citenamefont {Gr\o{}nbech-Jensen},\
  and\ \citenamefont {Nori}}]{Reichhardt01}%
  \BibitemOpen
  \bibfield  {author} {\bibinfo {author} {\bibfnamefont {C.}~\bibnamefont
  {Reichhardt}}, \bibinfo {author} {\bibfnamefont {C.~J.}\ \bibnamefont
  {Olson}}, \bibinfo {author} {\bibfnamefont {N.}~\bibnamefont
  {Gr\o{}nbech-Jensen}}, \ and\ \bibinfo {author} {\bibfnamefont
  {F.}~\bibnamefont {Nori}},\ }\bibfield  {title} {\enquote {\bibinfo {title}
  {Moving {Wigner} glasses and smectics: Dynamics of disordered {Wigner}
  crystals},}\ }\href {\doibase 10.1103/PhysRevLett.86.4354} {\bibfield
  {journal} {\bibinfo  {journal} {Phys. Rev. Lett.}\ }\textbf {\bibinfo
  {volume} {86}},\ \bibinfo {pages} {4354--4357} (\bibinfo {year}
  {2001})}\BibitemShut {NoStop}%
\bibitem [{\citenamefont {Reichhardt}\ and\ \citenamefont
  {Olson~Reichhardt}(2004)}]{Reichhardt04}%
  \BibitemOpen
  \bibfield  {author} {\bibinfo {author} {\bibfnamefont {C.}~\bibnamefont
  {Reichhardt}}\ and\ \bibinfo {author} {\bibfnamefont {C.~J.}\ \bibnamefont
  {Olson~Reichhardt}},\ }\bibfield  {title} {\enquote {\bibinfo {title} {Noise
  at the crossover from {W}igner liquid to {W}igner glass},}\ }\href {\doibase
  10.1103/PhysRevLett.93.176405} {\bibfield  {journal} {\bibinfo  {journal}
  {Phys. Rev. Lett.}\ }\textbf {\bibinfo {volume} {93}},\ \bibinfo {pages}
  {176405} (\bibinfo {year} {2004})}\BibitemShut {NoStop}%
\bibitem [{\citenamefont {Reichhardt}\ and\ \citenamefont
  {Reichhardt}(2021)}]{Reichhardt21}%
  \BibitemOpen
  \bibfield  {author} {\bibinfo {author} {\bibfnamefont {C.}~\bibnamefont
  {Reichhardt}}\ and\ \bibinfo {author} {\bibfnamefont {C.~J.~O.}\ \bibnamefont
  {Reichhardt}},\ }\bibfield  {title} {\enquote {\bibinfo {title} {Drive
  dependence of the {H}all angle for a sliding {W}igner crystal in a magnetic
  field},}\ }\href {\doibase 10.1103/PhysRevB.103.125107} {\bibfield  {journal}
  {\bibinfo  {journal} {Phys. Rev. B}\ }\textbf {\bibinfo {volume} {103}},\
  \bibinfo {pages} {125107} (\bibinfo {year} {2021})}\BibitemShut {NoStop}%
\bibitem [{\citenamefont {Reichhardt}\ and\ \citenamefont
  {Reichhardt}()}]{Reichhardt22}%
  \BibitemOpen
  \bibfield  {author} {\bibinfo {author} {\bibfnamefont {C.}~\bibnamefont
  {Reichhardt}}\ and\ \bibinfo {author} {\bibfnamefont {C.~J.~O.}\ \bibnamefont
  {Reichhardt}},\ }\bibfield  {title} {\enquote {\bibinfo {title} {Nonlinear
  dynamics, avalanches and noise for driven {W}igner crystals},}\ }\href@noop
  {} {\ }\Eprint {http://arxiv.org/abs/arXiv:2208.02929} {arXiv:2208.02929}
  \BibitemShut {NoStop}%
\bibitem [{\citenamefont {Lekner}(1991)}]{Lekner91}%
  \BibitemOpen
  \bibfield  {author} {\bibinfo {author} {\bibfnamefont {J.}~\bibnamefont
  {Lekner}},\ }\bibfield  {title} {\enquote {\bibinfo {title} {Summation of
  {C}oulomb fields in computer-simulated disordered-systems},}\ }\href
  {\doibase 10.1016/0378-4371(91)90226-3} {\bibfield  {journal} {\bibinfo
  {journal} {Physica A}\ }\textbf {\bibinfo {volume} {176}},\ \bibinfo {pages}
  {485--498} (\bibinfo {year} {1991})}\BibitemShut {NoStop}%
\bibitem [{\citenamefont {Gr{\o}nbech-Jensen}(1997)}]{GronbechJensen97a}%
  \BibitemOpen
  \bibfield  {author} {\bibinfo {author} {\bibfnamefont {N.}~\bibnamefont
  {Gr{\o}nbech-Jensen}},\ }\bibfield  {title} {\enquote {\bibinfo {title}
  {Lekner summation of long range interactions in periodic systems},}\ }\href
  {\doibase 10.1142/S0129183197001144} {\bibfield  {journal} {\bibinfo
  {journal} {Int. J. Mod. Phys. C}\ }\textbf {\bibinfo {volume} {8}},\ \bibinfo
  {pages} {1287--1297} (\bibinfo {year} {1997})}\BibitemShut {NoStop}%
\bibitem [{\citenamefont {Cha}\ and\ \citenamefont {Fertig}(1994)}]{Cha94a}%
  \BibitemOpen
  \bibfield  {author} {\bibinfo {author} {\bibfnamefont {M.-C.}\ \bibnamefont
  {Cha}}\ and\ \bibinfo {author} {\bibfnamefont {H.~A.}\ \bibnamefont
  {Fertig}},\ }\bibfield  {title} {\enquote {\bibinfo {title} {Topological
  defects, orientational order, and depinning of the electron solid in a random
  potential},}\ }\href {\doibase 10.1103/PhysRevB.50.14368} {\bibfield
  {journal} {\bibinfo  {journal} {Phys. Rev. B}\ }\textbf {\bibinfo {volume}
  {50}},\ \bibinfo {pages} {14368--14380} (\bibinfo {year} {1994})}\BibitemShut
  {NoStop}%
\end{thebibliography}%

\end{document}